\documentclass{emulateapj}

\usepackage{graphicx}
\usepackage[hypertex]{hyperref}

\def \eg {e.g.}
\def \ie {i.e.}
\def\spose#1{\hbox to 0pt{#1\hss}}
\def\ltsim{$\mathrel{\spose{\lower 3pt\hbox{$\sim$}}
        \raise 2.0pt\hbox{$<$}}$\thinspace}
\def\gtsim{$\mathrel{\spose{\lower 3pt\hbox{$\sim$}}
        \raise 2.0pt\hbox{$>$}}$\thinspace}
\newcommand{\thin }{\thinspace}

\newcommand{\lk }{${\rm L_K}$}

\newcommand{\msun }{${\rm M_{\odot}}$}
\newcommand{\lsun }{${\rm L_{\odot}}$}

\newcommand{\ergpscm}{${\rm erg\ s^{-1}\ cm^{-2}}$}
\newcommand{\mvir}{${\rm M_{vir}}$}
\newcommand{\rvir}{${\rm R_{vir}}$}

\newcommand{\mfive}{${\rm M_{500}}$}
\newcommand{\rfive}{${\rm R_{500}}$}
\newcommand{\rtwo}{${\rm R_{200}}$}
\newcommand{\rone}{${\rm R_{100}}$}

\newcommand{\rtwentyfive}{${\rm R_{2500}}$}
\newcommand{\cvir}{${\rm c_{vir}}$}

\newcommand{\reff}{${\rm R_{e}}$}

\newcommand{\src }{RXJ\thinspace 1159+5531}
\newcommand{\zfe }{${\rm Z_{Fe}}$}

\newcommand{\suzaku}{{\em Suzaku}}
\newcommand{\chandra }{{\em Chandra}}

\newcommand{\xspec }{{\em Xspec}}
\newcommand{\acis }{{\em ACIS}}

\newcommand{\minuit}{MINUIT}
\newcommand{\ciao }{{\em CIAO}}
\newcommand{\caldb }{{\em Caldb}}
\newcommand{\heasoft }{{\em Heasoft}}
\newcommand{\ned}{{\em{NED}}}

\newcommand{\xmm }{{\em XMM}}

\newcommand{\rosat }{{\em Rosat}}

\newcommand{\fb}{${\rm f_b}$}

\newcommand{\mtwentyfive}{${\rm M_{2500}}$}

\newcommand{\fgas}{${\rm f_{gas}}$}

\newcommand{\twomass}{2MASS}

\defcitealias{humphrey06a}{H06}
\defcitealias{humphrey08a}{H08}
\defcitealias{humphrey11a}{H11}

\slugcomment{Accepted for publication in The Astrophysical Journal}

\begin{document}
\title{Tracing the gas to the virial radius (\rone) in a fossil group}
\author{Philip J. Humphrey\altaffilmark{1}, David A. Buote\altaffilmark{1}, 
Fabrizio Brighenti\altaffilmark{2,3}, H\'el\`ene M.~L.~G. Flohic\altaffilmark{1,4}, Fabio Gastaldello\altaffilmark{1,5}, 
 William G. Mathews\altaffilmark{3} }
\altaffiltext{1}{Department of Physics and Astronomy, University of California at Irvine, 4129 Frederick Reines Hall, Irvine, CA 92697}
\altaffiltext{2}{Dipartimento di Astronomia, Universit\`a di Bologna, via Ranzani 1, Bologna 40127, Italy}
\altaffiltext{3}{University of California Observatories/Lick Observatory, University of California, Santa Cruz, CA 95064, USA}
\altaffiltext{4}{Departamento de Astronom\'{\i}a, Universidad de Chile, Casilla 36D Santiago, Chile}
\altaffiltext{5}{INAF, IASF, via Bassini 15, 20133 Milano, Italy}
%\altaffiltext{6}{Occhialini Fellow}
\begin{abstract}
We present a \chandra, \suzaku\ and \rosat\ study of the hot Intra Group Medium (IGrM) 
of the 
relaxed fossil group/ poor cluster \src. This group
exhibits an advantageous combination of flat surface brightness profile,
high luminosity and optimal distance, allowing  the gas to be
detected out to the virial radius (\rvir$\equiv R_{108}=$1100 kpc) in a single 
Suzaku pointing, while the complementary Chandra data reveal a round
morphology and relaxed IGrM image down to ~kpc scales.
We measure the IGrM entropy profile over $\sim$3 orders of magnitude in
radius, including 3 data bins beyond $\sim 0.5$\rtwo\ that 
have good azimuthal coverage ($>$30\%). We find no evidence that the 
profile flattens at large scales ($>$\rfive), and when corrected for 
the enclosed gas fraction, the entropy profile is very 
close to the predictions from self-similar structure formation 
simulations, as seen in massive clusters. Within \rvir, we 
measure a baryon fraction of $0.17\pm0.02$,
consistent with the Cosmological value. These results are in sharp
contrast to the gas behaviour at large scales recently reported in the Virgo and
Perseus clusters, and indicate that substantial gas clumping cannot be 
ubiquitous near \rvir, at least in highly evolved (fossil) groups.
\end{abstract}
\keywords{Cosmology: dark matter--- Xrays: galaxies: clusters --- 
Galaxies: groups: individual: RXJ1159+5531--- Galaxies: ISM}

\defcitealias{gastaldello07a}{G07}
\defcitealias{humphrey08a}{H08}
\defcitealias{humphrey11a}{H11}

\section{Introduction}
%As lower-mass counterparts to massive galaxy clusters, galaxy groups are expected to be 
%more susceptible to the effects of nongravitational heating on their IGrM. 
%This should manifest itself as systematically more elevated entropy profiles,
%and smaller gas fractions (as gas is evacuated from the central parts of the halo)
%in lower-mass systems, both of which are observed at scales smaller than 
%$\sim$\rfive\ \citep[\eg][]{gastaldello07a,gastaldello07b,sun08a,pratt10a}.
%Clearly, the exact radial distributions of the entropy and \fgas\ profiles 
%are a potentially powerful way to understand the non-gravitational processes that 
%shape group formation. Recent \chandra\ and \xmm\ studies have revealed 
%profiles that tend to flatten outside $\sim$0.1\rfive\
%\citep[\eg][]{gastaldello07a,mahdavi05a,finoguenov07a,sun08a,johnson09b}, which may reflect 
%the interplay of AGN- and supernovae wind heating of the IGrM \citep{mccarthy10a}.
Galaxy groups (defined here as bound systems with virial masses,
\mvir, in 
the range $\sim 10^{13}$--$10^{14}$\msun)
are essential ingredients in  the assembly of structure within the 
Universe. 
Locally $\sim$30\%\ of galaxies are found in 
groups that are less massive than $10^{14}$\msun\ 
\citep{eke04a}.
% Group scale halos; What I did was to use the Evrard (2004) relation
% quoted in Voit's review to get sigma=206 to sigma=445 
% I used Eke's catalogue (FGS), considering z<0.15; and counted how
% many galaxies are in halos with sigma<445...
Systems of this mass range have been implicated as key sites in both
efficient star formation \citep{springel03a}, and the 
morphological transformation of galaxies \citep{zabludoff98a}.
Typically the dominant baryonic component is, however, an extended,
hot gas halo, that can contain \gtsim70\%\ of the baryons at 
\mfive\gtsim$5\times 10^{13}$\msun\ \citep{giodini09a}\footnote{We define 
$M_\Delta$ as the mass within $R_\Delta$, \ie\ the geometrical 
radius within which the mean mass density of 
the system is $\Delta$ times the critical density of the Universe.}.
Such a massive hot gas halo can make them both luminous X-ray sources,
and, potentially, targets for detection in large numbers by future
Sunyaev-Zeldovich surveys \citep{haiman01a}.

X-ray observations provide the best means for studying
emission from the hot gas in groups. Of particular interest is the entropy proxy, 
$S=n_e^{-2/3} kT$ (where $n_e$ is the 
electron number density, $k$ is Boltzmann's constant and T is the
temperature), which is related to the specific entropy through a logarithm
and constant offset. If purely gravitational processes shape the 
energetics of the gas, self-similar structure formation simulations
predict an entropy profile that rises with
radius, r, such that
$S\propto T_0 r^{1.1}$, where $T_0$ is a characteristic temperature
\citep{tozzi01a,voit05b,kaiser86a}. Observations of galaxy clusters in fact reveal 
lower-mass objects to be systematically offset upwards from this relation,
but approaching it by $\sim$\rfive\ \citep[\eg][]{ponman03a,pratt10a}.
This indicates nongravitational energy injection,
 likely a consequence of preheating of the gas in filaments prior to 
accretion, or feedback due to star formation or a central AGN
\citep{voit03a,voit05a,ponman03a}. The fine balance between these
different processes should affect both the shape and normalization of the 
entropy proxy profile, making it a potentially powerful diagnostic 
\citep[\eg][]{borgani05a,mccarthy10a}, especially at the group scale,
where these effects should be comparatively more important.
Recent \chandra\ and \xmm\ observations of relaxed groups 
have revealed complex entropy profiles that flatten at large 
(\gtsim 0.1\rfive) and small scales, albeit with some scatter
\citep{gastaldello07b,mahdavi05a,finoguenov07a,humphrey08a,sun08a,cavagnolo09a,johnson09b,flohic11a}, possibly indicating that both star formation and AGN
feedback are important \citep{voit05a,mccarthy10a}.

%In a study of  31 galaxy clusters, \citet{pratt10a}
%found that the entropy proxy profiles out to 
%$\sim$\rfive, 
%could be brought into reasonable agreement with the self-similar
%predictions by scaling them with $\left( f_{gas}/ f_{b,U} \right)^{2/3}$,
%where \fgas\ is the enclosed gas fraction, and $f_{b,U}$ is the 
%comological baryon fraction \citep[0.17:][]{dunkley09a},
%suggesting that the primary impact of feedback is to redistribute the
%gas within the potential well (or eject it), rather than raise its temperature
%\citep{mathews11a}.
%This trend has now been confirmed in the central regions 
%(\ltsim 0.2--0.5\rfive) of galaxy groups \citep{flohic11a} and even
%in an isolated, Milky Way-mass elliptical galaxy 
%\citep[within \gtsim \rtwentyfive:][hereafter \citetalias{humphrey11a}]{humphrey11a}. 

Feedback also shapes the enclosed gas fraction (\fgas) profiles of 
groups and clusters, which are found to rise
with radius (\citealt{allen02a}; \citealt{vikhlinin06b}; \citetalias{gastaldello07a}; \citealt{sun08a}), but with groups containing a systematically
smaller fraction of their gas at small scales
(\eg\ Fig~11 of \citealt{humphrey11a}, hereafter \citetalias{humphrey11a}). 
Assuming they can be
converted into a reliable estimate for the total (initial) 
baryon fraction in the cluster
(\fb), \fgas\ measurements are a powerful 
cosmological tool, used either directly \citep{white93a,allen02a},
or in employing the gas mass as a virial mass proxy 
\citep[\eg][]{voevodkin04a}.
Massive clusters are preferred in this analysis
since \fgas\ measured at these scales should be closer to 
\fb\ than in poor clusters or groups.  Still, systems with masses
as low as  $\sim 2\times 10^{14}$\msun\ \citep{allen08a}, 
or even lower \citep{voevodkin04a} are routinely used.
Translating into \fb\ the \fgas\ values 
measured at scales far smaller than the virial
radius, \rvir\ (typically at $\sim$\rtwentyfive), involves a number of assumptions
that have yet to be robustly verified, especially in lower mass 
halos \citep[\eg][]{arnaud05b}. Intriguingly, for most of the group-scale
objects studied by \citet[][hereafter \citetalias{gastaldello07a}]{gastaldello07a}, 
extrapolating \fgas\
outside the field of view yielded global \fb\ constraints consistent
with the Universal baryon fraction \citep[0.17:][]{dunkley09a,komatsu11a},
in accord with the idea that 
 X-ray bright groups are baryonically closed
\citep{mathews05a}. Nevertheless, this extrapolation was subject to 
significant systematic uncertainty.

To date, most observations of the gas in groups and clusters have
been restricted to within $\sim$\rfive, and typically much
smaller scales are attained. 
For example, \rfive\ was only reached for 11 of the 
43 systems studied in the current largest group sample \citep{sun08a}.
In groups at these scales, \fgas\ is still only $\sim$0.07, in contrast to
$\sim$0.11 for clusters. Given the low, stable background for the \suzaku\
XIS instrument, recent work has begun to push measurements of the ICM 
in clusters out to $\sim$\rtwo, or beyond, but a coherent picture has
not yet emerged. While some studies have found consistency with model
predictions \citep{reiprich09a,hoshino10a}, deviations from
hydrostatic equilibrium \citep{bautz09a}, temperature asymmetries associated 
with large-scale structure  \citep{kawaharada10a} and, in 
three  systems, an unexpected flattening of the entropy proxy
profile outside $\sim$\rfive\ (PKS0745-191: \citealt{george09a}; 
Perseus: \citealt{simionescu11a}; Virgo: \citealt{urban11a}),
have been seen. \citeauthor{simionescu11a} and \citeauthor{urban11a} 
\citep[see also][]{nagai11a}
attributed the entropy flattening,
and associated over-estimate of \fgas, to putative clumpiness of the ICM in
the outskirts of the cluster, leading to a systematically biased gas density measurement.

Given the lack of a consistent story in the outskirts of clusters, the 
ubiquity of a clumpy ICM remains to be determined. The azimuthal
temperature variations in Perseus \citep{simionescu11a} and the large-scale
asymmetries in the X-ray image of Virgo \citep{bohringer94a} indicate
that these systems are not relaxed at large scales, consistent with
ongoing formation. This could give
rise to deviations from sphericity (complicating the deprojection) or
local hydrostatic equilibrium (hence an underestimate
of the gravitating mass, and an over-estimate of \fgas), and local distortions
in the entropy profile. On account of the proximity of these clusters, only a 
small fraction of the outer annuli were imaged in these studies, so it 
is unclear whether such large effects would be seen in azimuthally averaged
profiles. An additional concern 
is the need to minimize sources of systematic uncertainty in this 
background-dominated regime (\eg, \citealt{reiprich09a}, \citealt{bautz09a}, \citetalias{humphrey11a}).
\citet{eckert11a} demonstrated that the \rosat\ PSPC surface brightness profile
of PKS0745-191 disagrees at 7.7-$\sigma$ with the \suzaku\ density profile
inferred by \citet{george09a}. 
They attributed the discrepancy to systematic 
errors in the \citeauthor{george09a} background treatment. 
In the outermost bins, the deprojection procedure adopted by \citet{urban11a} 
and \citet{simionescu11a} depends sensitively on the correct modelling of 
projected emission from regions outside the field of view.
Furthermore, while \citeauthor{simionescu11a} attempted to mitigate the 
scattered light contamination from the cluster core, it is unclear 
how sensitive their results were to this correction. Similarly, the XMM-Newton profiles
of \citeauthor{urban11a} were sensitive to the treatment of the background.

In this paper, we present a joint \chandra\ and \suzaku\ study 
(carefully cross-checked with the archival \rosat\ data)
of the very relaxed galaxy group \src, allowing, for the first time
in a system with a mass as low as $\sim 10^{14}$\msun, the gas to be traced 
to scales as large as the virial radius, 
$R_{108}$. A single, giant elliptical galaxy dominates the 
stellar light, making it a prototypical ``fossil group''  \citep{vikhlinin99a}
and implying a highly evolved (\ie\ relaxed) dynamical state \citep{ponman94}.
The optimal combination of distance ($z=0.081$), mass (\rvir$=12$\arcmin)
and high surface brightness (such that gas is already detected to 
$\sim$\rfive\ with \chandra: \citealt{vikhlinin06b}; 
\citetalias{gastaldello07a}; \citealt{sun08a}), make it possible to 
measure the gas to \rvir\ in a single, modest (85~ks) \suzaku\ pointing.

The group was observed with the X-ray centroid slightly offset (by 5\arcmin)
from the \suzaku\ optical axis, to enable coverage out to $\sim$12\arcmin, 
while minimizing the complications of stray-light from having the 
X-ray peak outside the field of view. Although this configuration
limited the number of radial bins we can study, by combining the 
\chandra\ and \suzaku\ data, we were able to  
achieve $\sim$3 spatial bins outside $\sim 0.5$\rtwo, with \gtsim 30\%\ 
azimuthal coverage, 
in comparison with $\sim$6 bins, and $\sim$15\%\ coverage in the
$\sim$3 times longer \suzaku\ exposure of Perseus \citep{simionescu11a}.
In conjunction with the archival \chandra\ data, we were able to measure the 
gas properties over almost three orders of magnitude in radius.

Previous studies of \src, based on the \chandra\ data
(\citealt{vikhlinin06b}; \citetalias{gastaldello07a}; 
\citealt{sun08a}),
have reported conflicting parameterizations of the 
gravitating mass profile.
Adopting the popular Navarro-Frenk-White 
\citep[NFW:][]{navarro97} profile (plus, in the case of
\citetalias{gastaldello07a}, a baryonic component), 
\mfive\ inferred from these studies
has varied  significantly 
from $\sim 6\times 10^{13}$--$10^{14}$\msun, and the corresponding 
NFW concentration parameter from $\sim$1.7--5.6,
which remains a puzzle (see 
\S~\ref{sect_mass_profile}). 
The addition of new data at large scales should help
pin down the mass profile more precisely, particularly if the 
scale radius were as high as found by \citeauthor{vikhlinin06b} (400~kpc).
In this paper, we employed the ``forward-fitting'' mass analysis 
techniques outlined in \citetalias{humphrey11a}, which enable finer
control of systematic uncertainties than more traditional methods
\citep{buote11a}.

 We assumed a flat cosmology with $H_0 =70 {\rm km\ s^{-1}}$
and $\Omega_\Lambda=0.7$. 
We adopted $R_{108}$ as the virial radius (\rvir), based on the approximation
of \citet{bryan98a} for the redshift of \src.
Unless otherwise stated, all error-bars represent 1-$\sigma$ confidence limits (which, for our 
Bayesian analysis, implies the marginalized region of parameter space within which the integrated 
probability is 68\%).

\section{Data Reduction and Analysis}
\begin{figure*}
\centering
\includegraphics[width=6in]{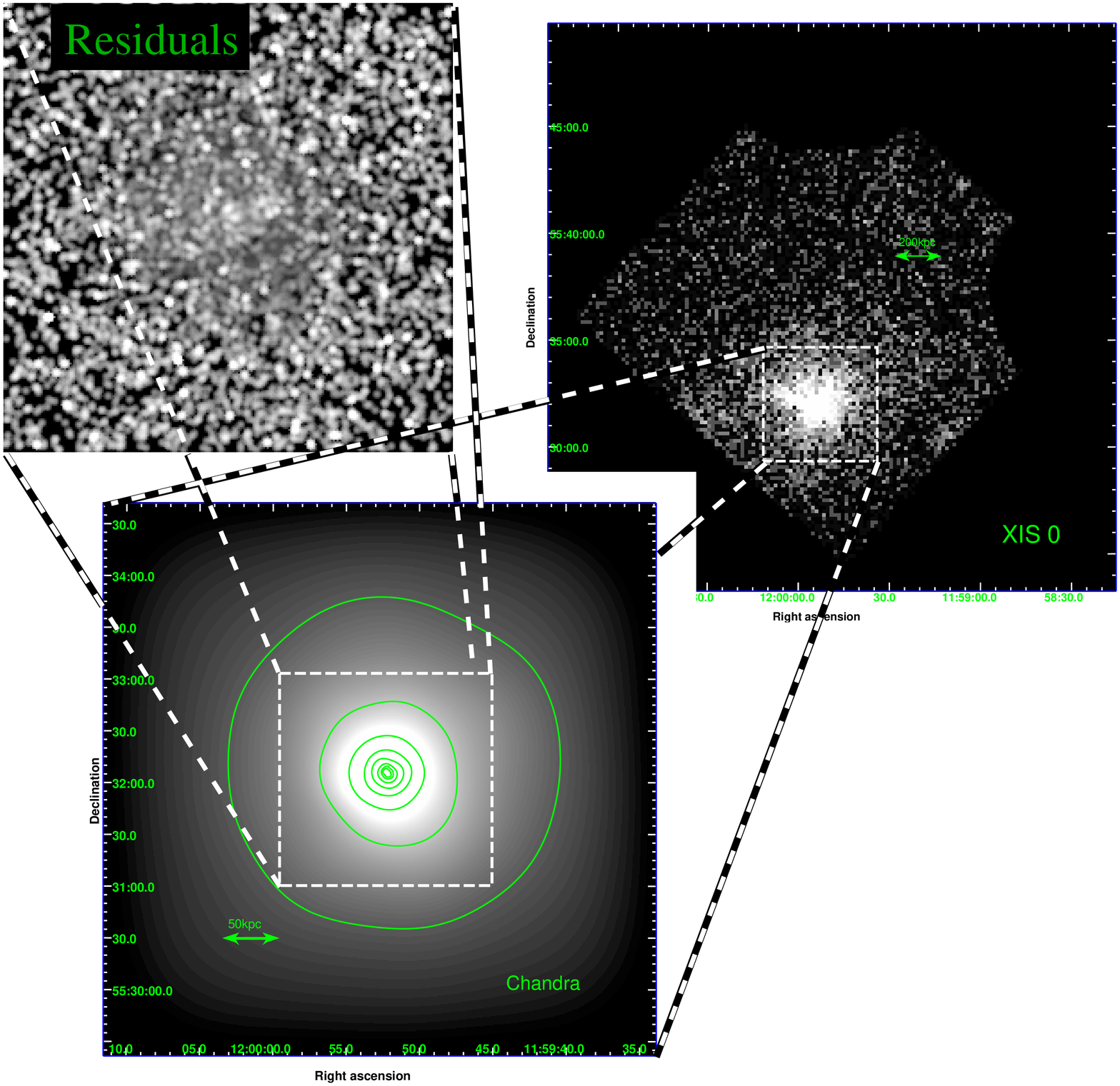}
\caption{\suzaku\ XIS0 (top right) and \chandra\ (bottom left) images of \src\ \label{fig_images}.
The apparent asymmetries in the XIS0 image are due to the asymmetric point spread function of
the telescope, rather than the intrinsic shape of the source. The \chandra\ image has been
cleaned of point sources and mildly smoothed. The smoothing scale varied from $\sim$1\arcsec\
at the smallest scales, to $\sim$1.1\arcmin\ (100~kpc) 
at the outer part of the image (see text). At the top left, we show a 
``residual significance'' image (see text) of the centre of the group, 
indicating deviations from a smooth model fit to the (\chandra) X-ray isophotes. There is no
obvious large-scale feature in this map, indicating the system is largely relaxed.
\label{fig_image}}
\end{figure*}
\begin{figure*}
\centering
\includegraphics[width=6in]{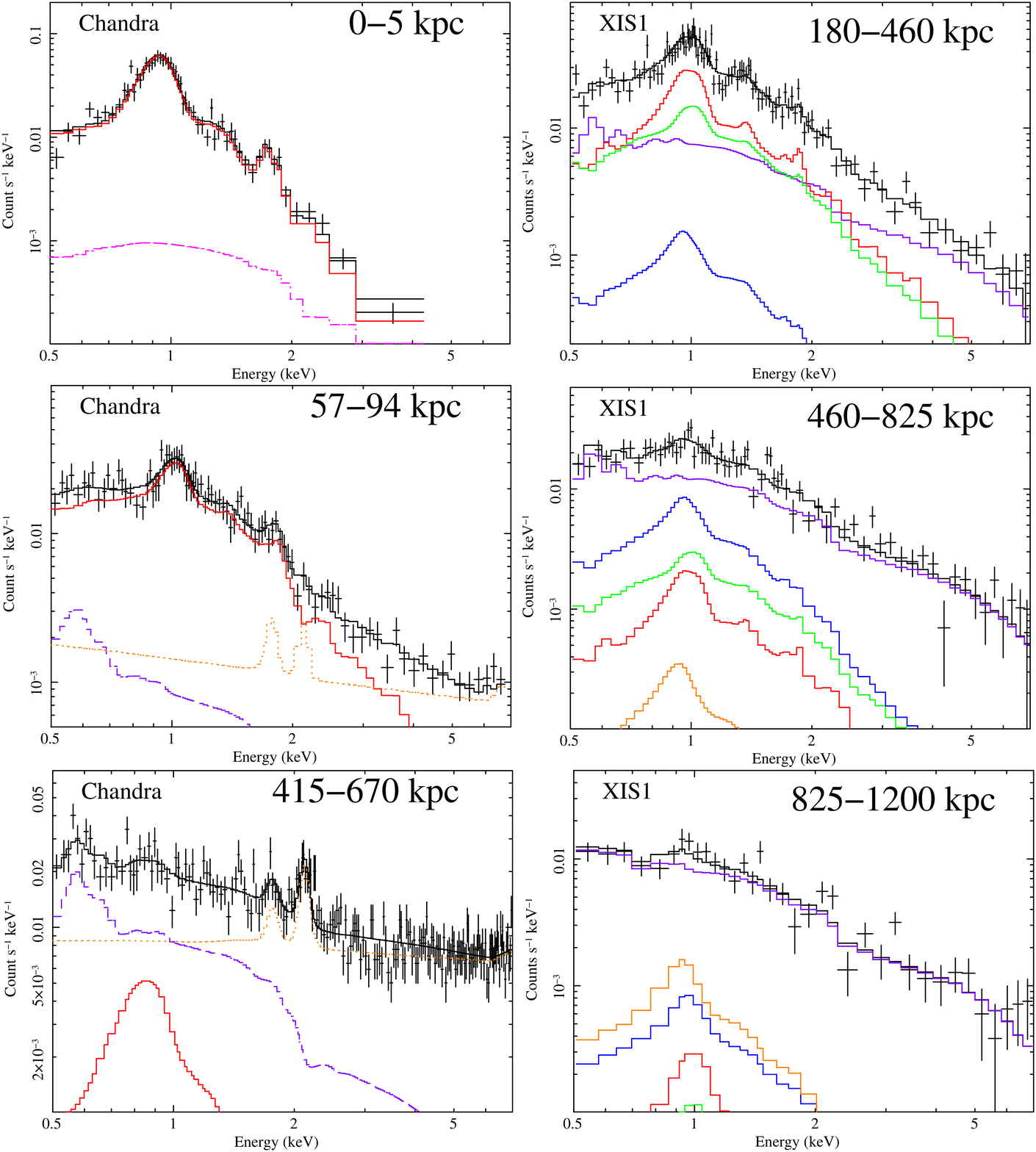}
\caption{Representative \chandra\ and \suzaku\ XIS1 spectra for \src, shown (\chandra) without 
background subtraction, or (\suzaku) with the instrumental background subtracted. 
In addition to the data, we show the best-fitting model, folded through the instrumental
response (solid black line), along with the decomposition of this model into its various
components. For the \chandra\ data, we show the hot gas contribution (solid red line), the 
composite emission from X-ray binaries (dash-dot magenta line), the instrumental background
(dotted orange) and the cosmic X-ray background (dashed purple line). For the XIS1 data, we show the 
hot gas emission from each annulus as solid lines (red, green, blue and orange indicating, 
respectively, the 0--2\arcmin, 2--5\arcmin, 5--9\arcmin and 9--13\arcmin\ apertures, respectively)
, sky background (using the same colour scheme as for 
\chandra). For \chandra, the  background is dominated by the instrumental component, 
whereas for \suzaku, which has a lower instrumental background but is less able to resolve
the cosmic component into individual point sources, the cosmic component dominates. 
\label{fig_spectra}}
\end{figure*}
\subsection{Chandra} \label{sect_chandra}
The region of sky containing \src\ was imaged by the \acis\ instrument aboard \chandra\
on two separate occasions. We consider here only the deep data taken in the ACIS-S configuration
(Observation ID 4964; beginning on Feb 11 2004). A shallower ACIS-I observation
was also available, but to simplify the analysis (in particular, the background modelling),
we chose not to include it in our study.
The data-reduction was carried out as
described in \citetalias{humphrey11a}, using the 
\ciao\thinspace 4.1 and \heasoft\thinspace 6.8 software suites, in 
conjunction with the \chandra\ calibration database (\caldb) version 4.1.2. 
Briefly, the data were reprocessed from the ``level 1'' events files, following the
standard data reduction threads\footnote{http://cxc.harvard.edu/ciao/threads/index.html}. 
Periods of high background were identified by eye in the lightcurve from a low surface-brightness
region of the CCDs and data from these intervals were excised, leaving a total exposure
of 75~ks. Point sources were 
detected in the 0.3--7.0~keV image with the {\tt wavdetect} \ciao\ task, which was 
supplied a 1.7~keV exposure map to minimize spurious detections at chip boundaries. The 
detection threshold ($10^{-6}$) guaranteed \ltsim 1 spurious source per CCD. 
All detected sources were confirmed visually, and appropriate elliptical regions containing 
$\sim99$\%\ of the source photons were generated. 

In Fig~\ref{fig_image}, we show a 
smoothed, flat-fielded \chandra\ image, having removed the point sources with the 
algorithm outlined in \citet{fang09a}. The image was smoothed with a Gaussian kernel, the 
width of which varied with distance from the nominal X-ray centroid 
according to an arbitrary powerlaw, ranging from $\sim$1\arcsec\ at the centre of the 
image to $\sim$1\arcmin\ at its edge. The image is smooth and very round, consistent with 
the relaxed morphology expected for a fossil system. To search for more subtle structure,
we used dedicated software to fit an elliptical beta model (with constant ellipticity)
to the central 2\arcmin-wide portion of the unsmoothed (flat-fielded) image\footnote{We
obtained a best-fitting major axis core radius of 2\arcsec, $\beta=$0.51 and axis ratio
of 0.9.}. In Fig~\ref{fig_image}, we plot $(data - model)^2/model$, corresponding 
(approximately) to the $\chi^2$ residuals from this fit. To bring out the structure,
we smoothed this image with a Gaussian kernel of width 3~pixels. There is weak evidence
of a small ($\sim$4\arcsec), coherent structure in the residuals within the central 
($\sim$10~kpc) region, which may imply a small depression in the surface brightness 
(by $\sim$20--30\%). The formal significance of this feature depends sensitively on 
prior information \citep[in particular, the region of the image over which one searches
for structures; \eg][]{kaastra06a}. Still, even if the feature is real, it should not 
give rise to a significant error in the recovered, azimuthally averaged mass profile
\citep[see][]{buote11a,churazov08a}. Aside from this modest feature, the overall 
lack of significant, coherent residuals indicates that the X-ray image is very relaxed.

Spectra were extracted in a series of contiguous, concentric annuli centred at the 
X-ray centroid. The widths of the annuli were chosen to contain approximately the 
same number of background-subtracted counts, while ensuring sufficient photons for
useful spectral analysis. The resulting annuli had widths larger than $\sim$3\arcsec,
which is sufficient to prevent spectral mixing between adjacent annuli on account
of the finite spatial resolution of the mirrors. Data in the vicinity of point sources
and chip gaps were excluded. We extracted spectra from all the active chips (excluding
S4, which suffers from noise). Appropriate count-weighted spectral response matrices
were generated for each annulus with the standard \ciao\ tools {\tt mkwarf} and 
{\tt mkacisrmf}. Representative spectra, without background subtraction, are
shown in Fig~\ref{fig_spectra}.

Spectral-fitting was carried out in the energy band 0.5--7.0~keV, using \xspec\ 
vers.\ 12.5.1n, by minimizing the C-statistic to mitigate biases that arise 
(even in the high-count regime) when using the standard $\chi^2$ approximations for 
Poisson-distributed data \citep{humphrey09b}. To aid convergence, we rebinned 
the spectra to ensure at least 20 photons per bin. The data in all annuli were
fitted simultaneously, to enable  the source and background components to be 
modelled at the same time. In keeping with \citetalias{humphrey11a}
and \citet{gastaldello07a}, we modelled the {\em projected} (rather than
the {\em deprojected}) source emission in each annulus as 
coming from a single APEC plasma model with variable abundances, modified by 
foreground Galactic absorption \citep{dickey90}\footnote{We discuss the results from
a {\em deprojected} analysis in \S~\ref{sect_syserr_3d}.}. We allowed the total abundance 
of Fe (\zfe) and the abundance {\em ratios} with respect to Fe of the elements 
O, Ne, Mg, Si and Ni to vary in each annulus. The abundance of He, and the 
abundance {\em ratios} of the other elements were fixed to 1 Solar \citep{asplund04a}.
To improve S/N, we tied \zfe\ between the outermost two annuli, and constrained 
the abundance ratios to be the same in all annuli.

To account for emission from undetected LMXBs in the central galaxy, we included an 
additional 7.3~keV bremsstrahlung component. Since the number of X-ray point sources 
is approximately proportional to the stellar light \citep[\eg][]{humphrey08b}, the 
relative normalization of this component between each
annulus was fixed to match the relative K-band luminosity in the matching regions, which we 
measured from the \twomass\ image. This component is only important in the very central region
(\ltsim 20~kpc) of the system.

To accommodate the background, we included additional spectral components in our 
fits. Specifically, we included two (unabsorbed)
APEC components (kT=0.07~keV and 0.2~keV) and an
(absorbed) powerlaw component ($\Gamma=1.41$; \citealt{deluca04a}). The normalization of each component within each annulus
was assumed to scale with the extraction area, but the total normalizations were fitted freely. 
We discuss the possible impact of an additional, ``Solar wind charge exchange'' background component
in \S~\ref{sect_syserr_swcx}.
To account for the instrumental background, we included a number of Gaussian lines and a 
broken powerlaw model,
which were not folded through the ARF. We included separate instrumental components for the front- and 
back-illuminated chips, and assumed that the normalization of each component scaled with the area of the 
extraction annulus which overlapped the appropriate chips. The normalization of 
each component, and the shape
of the instrumental components, were allowed to fit freely.
We included two Gaussian lines (at 1.77 and $\sim$2.2~keV), the 
intrinsic widths of which were fixed to zero. The energies of the $\sim$2.2~keV lines were allowed to 
fit freely, as were as the normalizations of all the components.
To verify the fit had not become trapped in a local minimum,
we explored the local parameter space by stepping individual parameters over
a range centred around the best-fitting value 
\citep[analogous to computing error-bars with the algorithm of][]{cash76}.
The covariance matrix (which contains the error bars) was computed via the efficient Monte Carlo technique
outlined in \citet{humphrey06a}, and we carried out 250 error simulations.

The best-fitting models are shown in Fig~\ref{fig_spectra} for a representative 
selection of spectra. While the instrumental background is clearly significant at 
high energies (\gtsim 2~keV), only in the outermost annulus does the source signal 
fall below the background level. Nevertheless, given the optimal temperature of the 
gas ($\simeq$1~keV), the Fe L-shell peak is still visible as a small ``bump'' in the 
spectrum at $\sim$0.9~keV, enabling the gas temperature and density to be constrained
(this is similar to the outermost annuli of NGC\thin 5044, studied by \citealt{buote04c}).

\subsection{Suzaku}
The region of sky containing \src\ was imaged by \suzaku\ beginning on Feb 5 2009
(observation ID 804051010), with three of the XIS units operating. Data-reduction was 
performed using the \heasoft\ 6.8
software suite, in conjunction with the XIS calibration database (\caldb) version 20090925.
To ensure up-to-date calibration, the unscreened data were re-pipelined with the {\tt aepipeline} task
and analysed following 
the standard data-reduction guidelines\footnote{{http://heasarc.gsfc.nasa.gov/docs/suzaku/analysis/abc/}}. 
Since the data for each instrument were divided into differently telemetered events file
formats, we converted the ``$5\times5$'' formatted data into  ``$3\times3$'' format,
and merged them with the ``$3\times3$'' events files.
The lightcurve of each instrument
was examined for periods of anomalously high background, but no significant amount of data
was found to be contaminated in this way, leaving 85~ks of total ``cleaned'' exposure time.
85~ks total exposure time survived flare cleaning. In Fig~\ref{fig_images}, we show
the 0.5--7.0~keV image for the XIS0 detector, excluding data  in the vicinity of the 
calibration sources. By visual inspection of the images for all three active detectors,
we found only 1 bright point-source (coinciding with one of the calibration sources in the 
XIS0 image). In subsequent analysis, we excluded a circular region of 2.5\arcmin\ radius,
centred on this source (which should eliminate \gtsim 90\%\ of the source photons from
contaminating any spectra).

Since \src\ was slightly offset from the centre of the field of view, it was possible to 
extract spectra out to scales of $\sim$13\arcmin\ from the single, pointed observation.
Therefore, spectra were extracted in four concentric annuli (0--2\arcmin, 2--5\arcmin,  5--9\arcmin\
and 9--13\arcmin), 
centred at the nominal position of \src\ in the field of view of each instrument.
Due to the large field of view of \suzaku, our spectral extraction regions actually
achieve $\sim$55--60\%\ azimuthal coverage in the 5--9\arcmin\ aperture, and $\sim$27\%\ 
in the 9--13\arcmin\ region.
Data in the vicinity of the calibration sources and the identified point sources were excluded.
For each spectrum, we generated an associated redistribution matrix file (RMF) using the {\tt xisrmfgen}
tool and an estimate of the instrumental background with the {\tt xisnxbgen}  task.
Ancillary response files (ARFs) were generated for each spectrum with the {\tt xissimarfgen}
tool \citep{ishisaki07a}, which models the telescope's optics through ray-tracing.
Since the point-spread function of the telescope is very large ($\sim$2\arcmin\ half power diameter), 
even with such large apertures it is necessary to account for spectral mixing between each annulus.
We did this by employing the algorithm described in \citetalias{humphrey11a}.

We modelled the spectrum in each annulus as the sum of source plus sky background components. The 
instrumental background (which is not the dominant background component; \citetalias{humphrey11a})
was subtracted directly, using the model generated with the {\tt xisnxbgen}  task.
For the source, we included an APEC component, for which the Fe abundance and 
abundance ratios of other species
were allowed to vary (as for the \chandra\ data). The abundance ratios were 
tied between all annuli, and \zfe\ was tied between the outermost two annuli. 
In the central bin, we also included
a 7.3~keV bremsstrahlung component, to account for possible LMXBs in the central galaxy.
For this component, we obtained a total X-ray luminosity per unit
K-band optical light of
$L_{X,LMXB}/L_K = 2.7\pm 3.9 \times 10^{29} erg\ s^{-1} L_\odot^{-1}$,
in good agreement with the \chandra\ result of 
$6.0\pm 1.9 \times 10^{29} erg\ s^{-1} L_\odot^{-1}$. 
This self-consistency gives us confidence in our 
treatment of the background (described below), and the lack of bright source
 contamination in the 
\suzaku\ field. This ratio is also marginally consistent with the mean scaling
determined from local galaxies \citep{humphrey08a}.

The \chandra\ data reveals a strong temperature gradient within the central 
$\sim$2\arcmin. Since we can generally approximate the integrated spectrum
of a region in which there is a temperature gradient by the sum of two 
APEC components \citep{buote99a,buote03a}, we added a second APEC component in the central bin.
Although we did not consider the fitted temperature or density in this bin 
in our subsequent analysis, it was still necessary to fit a reasonably 
accurate model to the data, as some fraction of the photons from this 
region are scattered into the surrounding annuli.
The spectra from all instruments were fitted simultaneously, allowing
additional multiplicative constants in the fit to enable the relative 
normalization of the model for each instrument to vary with respect to 
the XIS 0.

We found that this model was able to fit the data well\footnote{Since the
C-statistic is not easily interpretable as a goodness-of-fit, we 
also performed a $\chi^2$ fit. Although, in general, such fits are 
more biased \citep{humphrey09b}, the best-fitting parameters were 
very close to those obtained with the C-statistic, and the fit 
was formally very good ($\chi^2$/dof=1787/1811).}. 
We show in Fig~\ref{fig_spectra} representative spectra for the XIS1 
instrument, showing various model components. As is immediately 
clear, mixing between annuli is significant. Furthermore, the data
are clearly background-dominated in the outermost annulus (while the
source and background are comparable in the 6--10\arcmin\ region). 
Nevertheless, given the $\sim$keV temperature of the gas, the 
Fe L-shell peak is still clearly visible over the cosmic X-ray background
component (\S~\ref{sect_cxb}), and so the gas properties can be 
measured with reasonable precision.

\subsection{Rosat}
\src\ was observed serendipitously by \rosat\ in the outskirts 
($\sim$17\arcmin\ off-axis; $\sim$2.5\arcmin\ inside the support-ring
structure) of a  pointed PSPC observation (observation 
ID 700055n00). Given its relatively poor spatial resolution and 
limited energy pass-band, we only required to measure the \rosat\ surface
brightness profile, centred at the source (see \S~\ref{sect_rosat_sb} for more details).
We do not expect much extended emission from the 
principal target of the observation, an unrelated, foreground Sy1
galaxy (NGC\thin 3998). Therefore, we extracted the surface brightness
profile of \src\ out to scales of $\sim$25\arcmin\ ($\sim 2$\rvir). 
To prepare the data, we followed the standard \rosat\ data-reduction 
recipe \citep[see, for example][]{fang09a}. A full field of view
image with 15\arcsec\ pixel size was generated in the 0.42--2.0~keV band, 
and the corresponding
exposure maps were generated with the pcexpmap task. Point sources
were detected with the {\em wavdetect} \ciao\ task to search for 
structure at scales of 1, 2, 4 and 8 pixels, and supplied with the 
exposure map to minimize spurious detections at the image boundaries.
The detection threshold was set to $10^{-6}$, implying \ltsim 0.1
spurious detections per image, and all point sources were visually
confirmed. We used the {\em ao} and {\em castpart} \heasoft\ 
tasks to generate, respectively, the scattered Solar X-ray background
and the particle background contribution, 
which were then subtracted from the image.
The surface brightness profile was generated in 
concentric annuli from the flat-fielded image, excluding data 
from the vicinity of any detected point source and, to maximize the 
ability to identify and exclude point sources, all photons not inside
the shadow of the inner PSPC support ring. We discuss the surface brightness
analysis in detail in \S~\ref{sect_rosat_sb}.

\section{Spectral fitting results}
\subsection{Metal abundances} \label{sect_abund}
\begin{figure}
\centering
\includegraphics[height=3.3in,angle=270]{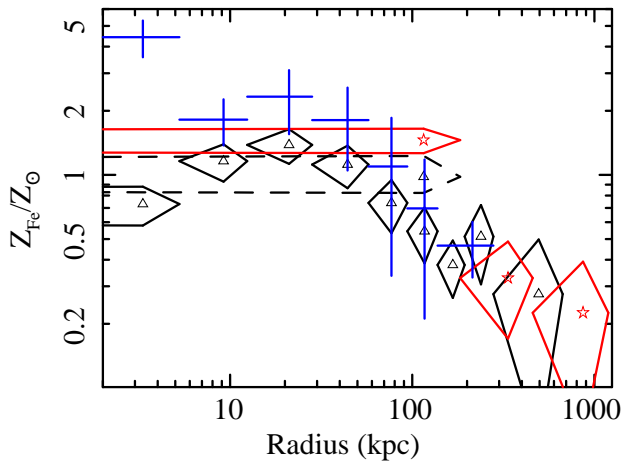}
\caption{Projected Fe abundance profile, measured with \chandra\ (triangles)
and \suzaku\ (stars). The dashed line indicates the \chandra\ abundance
measured in same region as the central \suzaku\ bin. Note the overall
good agreement between \chandra\ and \suzaku. We also overlay
(crosses) the deprojected abundance profile measured with \chandra,
including a second temperature component in the innermost bin. Note that
the central dip seen in the projected profile is absent in this case.
\label{fig_abund_profile}}
\end{figure}
\begin{figure}
\centering
\includegraphics[height=3.3in,angle=270]{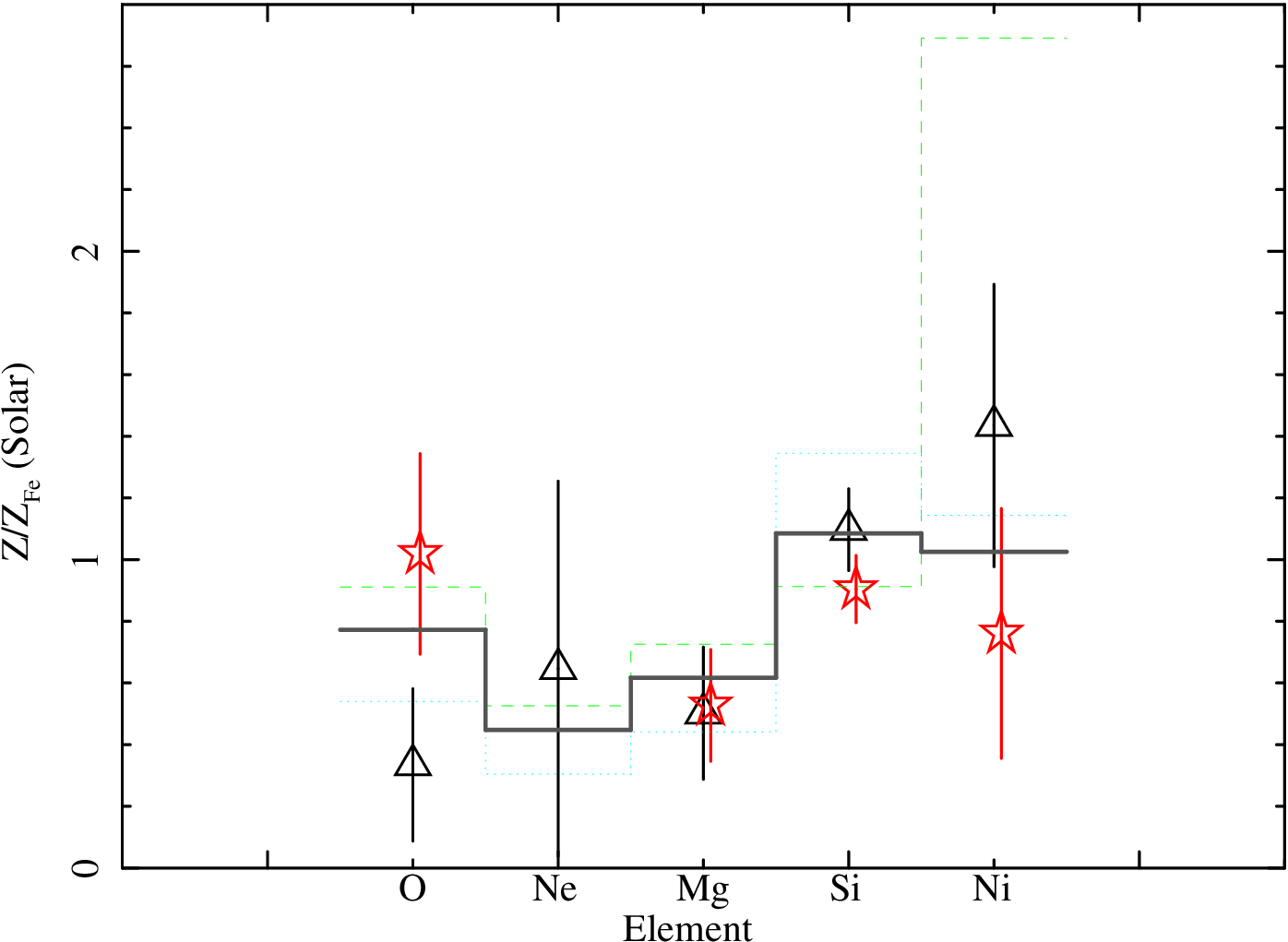}
\caption{Abundance ratios with respect to Fe, expressed relative to the 
Solar ratios \citep{asplund04a}. Chandra data are marked with triangles, and 
\suzaku\ data with stars. The solid line is the best-fit model 
($\chi^2$/dof=5.7/8) where the enrichment comes from SNIa and SNII, assuming
the WDD2 SNIa yields from \citet{nomoto97b} and the SNII yields from
\citet{nomoto97a}. The SNIa enrichment fraction is $0.82\pm0.05$. For reference,
we also show the best fits with the W7 (green dashed line) and WDD1 (blue dotted line)
SNIa yields, neither of which are formally acceptable.
\label{fig_abund_ratios}}
\end{figure}
In Fig~\ref{fig_abund_profile}, we  show the projected abundance profile,
which is centrally peaked, as expected for a relaxed galaxy group
and consistent with a picture in which metal enrichment
is facilitated by mass-loss from stars and supernovae in the 
central galaxy \citep{mathews03a}.  There is good agreement between the profiles measured
outside $\sim$100~kpc with \chandra\ and \suzaku, supporting our 
treatment of these data. To effect a comparison within the 
central 2\arcmin, where there is a strong abundance gradient
revealed by \chandra, but only a single \suzaku\ data-bin, we
extracted a single \chandra\ spectrum for this region, and 
fitted it with the same model as the \suzaku\ data. The best-fitting
\zfe\ (shown on Fig~\ref{fig_abund_profile}) was consistent (within
$\sim$1.5-$\sigma$) with the \suzaku\ measurement.
The good agreement over the entire radial range is encouraging,
and suggests that our treatment of the spectral mixing between
the \suzaku\ annuli is approximately correct.

We note the slight dip in the central \chandra\ 
bin, similar to features that have been reported in some other galaxy
groups and clusters. Still, if hot gas components with a range of 
different temperatures are found along the line-of-sight to the 
central annulus, the measured \zfe\ may be systematically 
under-estimated due to the ``Fe bias'' \citep{buote98c,buote00a}. Such a situation
could arise either due to projection effects, or if there is a strong
temperature gradient within that annulus. 
To investigate this, we carried out a 
deprojection analysis (see \citetalias{humphrey11a}, or \S~\ref{sect_syserr_3d} 
of this paper, for a detailed description of the deprojection procedure),
and, furthermore, included an additional APEC component within the 
cental bin, with abundances tied to those of the other component. 
Since the deprojection procedure tends to increase the error-bars
on the measured quantities, we found it necessary to fix the abundance 
outside $\sim$400~kpc to 0.2 Solar (consistent with the projected
results).
We found that the second hot gas component was formally required in
the central bin (the improvement in the C-statistic was 13.7 for a difference of 
2 degrees of freedom when it was added); the two temperatures in this 
bin were $1.29\pm0.23$~keV and $0.72\pm0.17$keV, consistent with a continuing 
temperature decline in the group's centre, although it may also 
reflect deficiencies in the deprojection procedure. 

The deprojected 
\zfe\ profile is shown in Fig~\ref{fig_abund_profile}, and it also exhibits a 
centrally-peaked shape and, significantly, no evidence of a 
central abundance dip. This strongly suggests that the feature
seen in the central bin of the projected data 
is simply an artefact of the Fe bias. 

In addition to the abundance profile, the data also provide
interesting constraints on the abundance {\em ratios} with 
respect to Fe of various species. These are summarized in 
Fig~\ref{fig_abund_ratios}; we find excellent agreement between
\chandra\ and \suzaku. Overlaid on this figure, we show the 
best-fitting abundance ratio patterns predicted by 
combining the metal yields for type Ia and type II supernovae,
through which Fe and the $\alpha$ elements are primarily processed.
We adopted the SNII yields from \citet{nomoto97a}, and explored
three different theoretical SNIa yields from \citet{nomoto97a},
specifically the so-called ``W7'', ``WDD1'' and ``WDD2'' models.
We found the WDD2 model fits the data well, provided $82\pm5$\%\
of the Fe was synthesized in type Ia supernovae. This is 
consistent with measurements in the central parts of other
galaxy groups and clusters 
(\citealt{humphrey05a}, and references therein; \citealt{werner08b},
for a review).

\subsection{Temperature and density profiles}
\begin{figure*}
\centering
\includegraphics[width=6.5in]{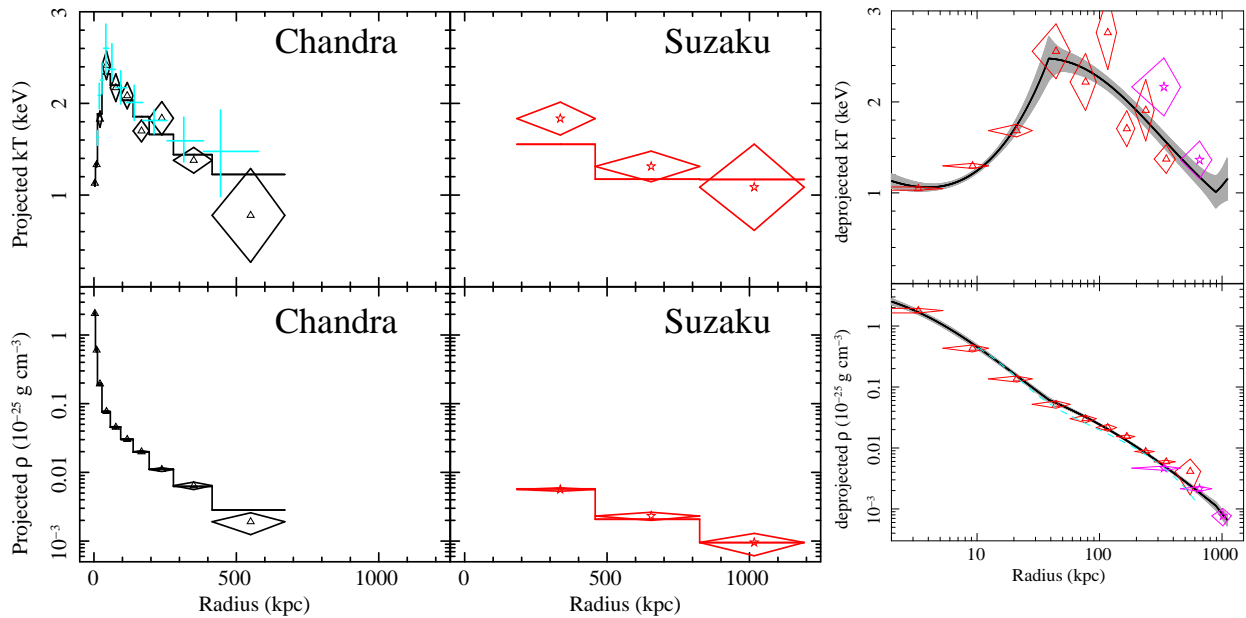}
\caption{Radial temperature (top panels) and density (bottom panels)
profiles for \src. We show the projected \chandra\ profiles in the left
column (triangles), the projected \suzaku\ profiles in the right column 
(stars), and the deprojected profiles in the right column. Overlaid are the 
best hydrostatic model fits to each dataset, which match the data very well
and, for the deprojected profiles, we show in grey the 1-$\sigma$
confidence region for the model.
We exclude the central \suzaku\ bin from the fitting (although it was used
for abundance determination) since two temperature components were required,
to account for the strong temperature gradient. In the upper left panel,
we show the projected temperature profile measured by \citet{vikhlinin06b}
from the \chandra\ data (light blue crosses), which agrees well with our 
data. In the bottom right panel, we show (light blue dashed line) 
the deprojected density profile found by \citet{vikhlinin06b}, which is 
also in reasonable agreement with our data. 
 \label{fig_kt_rho}}
\end{figure*}
The projected gas temperature and density\footnote{We define the 
projected density in any given annulus as the mean gas density if all 
the emission measured in the annulus originates from a region defined by 
the intersection of a cylindrical shell and a (concentric) 
spherical shell, both of 
which have the same inner and outer radii as the annulus.} profiles 
are shown in Fig~\ref{fig_kt_rho}
for \chandra\ and \suzaku. 
 These profiles span almost 3 orders of 
magnitude in radius. {The outermost \suzaku\ data-bin reaches 
$\sim$1200~kpc, which is $\sim$\rvir, as found by \citet{gastaldello07a}; see also 
\S~\ref{sect_mass_profile}}. There is excellent agreement between the 
results for the two satellites, giving us confidence in our treatment
of the mixing between \suzaku\ annuli, and our treatment of the background
components, which differ slightly between both satellites. Since two
APEC components were employed in the central \suzaku\ bin, we do not 
show the gas density or temperature for that bin. The \chandra\
temperature profile agrees well with that found by 
\citet{vikhlinin06b}. In Fig~\ref{fig_kt_rho}, we also show the 
deprojected temperature profile (see \S~\ref{sect_syserr_3d} for a description of how
this was obtained), which is in good agreement with the deprojected
profile reported for the \chandra\ data by \citet{gastaldello07a}.

As discussed in \S~\ref{sect_abund}, there is evidence that the 
gas temperature may continue to fall in the central bin, possibly
requiring a second APEC component to be added to the spectrum 
in this region. Nevertheless, the results for the single
temperature fit can still be interpreted in our mass-fitting
analysis, provided care is taken to compute an appropriately
weighted average gas density and temperature in that region
(see \citetalias{humphrey11a} for more details). In any case, we found that our 
results were relatively insensitive to the inclusion or omission
of that particular annulus (\S~\ref{sect_syserr_stars}).

\section{Mass modelling}
\subsection{Method} \label{sect_method}
We translated the density and temperature profiles into mass constraints using 
the entropy-based ``forward-fitting'' technique developed in our recent 
papers (\citealt{humphrey08a,humphrey09d}; \citetalias{humphrey11a}; 
\citealt{buote11a}, for a review).
Briefly, given parametrized profiles of 
``entropy'' (S=kT$n_e^{-2/3}$, where $n_e$ is the electron density) 
and gravitating mass (excluding the gas mass, which is 
computed self-consistently), plus the gas density at some canonical  radius
(for which we used 10~kpc),
the three-dimensional temperature and density profiles can be calculated, under the 
hydrostatic equilibrium approximation and fitted to the data. 
This model assumes spherical symmetry, which is a standard
assumption in X-ray hydrostatic modelling, even when the X-ray isophotes are not perfectly 
round, since deviations from sphericity likely only contribute a very small error
($\sim$few percent) into the recovered mass profile and baryon fraction
(\citealt{buote11c}; see also \citealt{piffaretti03a,gavazzi05a}).

For the gravitating mass model, we assumed an NFW \citep{navarro97} dark matter halo,
the virial mass and concentration of which were free fit parameters, plus a model
for the stellar mass. Since the projected stellar light of the central galaxy is known to
be well-fitted by a de Vaucouleurs model \citep{vikhlinin99a}, for the stellar
mass component, we adopted a deprojected de Vaucouleurs model, using the analytical
approximation of \citet{prugniel97a}. We fixed the effective radius and luminosity
of this component to the 
K-band values (9.8~kpc, and $1.0\times 10^{12}$\lsun\ at the distance to \src), 
inferred from \twomass\ \citep{gastaldello07a}, and allowed the K-band M/L ratio 
to be fitted freely. We additionally included a supermassive black hole,
with mass fixed at $2.4\times 10^9$\msun, based on the \lk\ of the galaxy
and the black-hole mass {\em versus} V-band bulge luminosity relation of
\citep{gultekin09a}, assuming 
$L_V/L_K=0.24$, which is typical for an old stellar population with Solar 
abundances\footnote{We note that the updated black hole mass {\em versus}
K-band luminosity relation of \citet[][their eqn~14]{graham07a} gives a slightly smaller mass for
the black hole ($1.8\times 10^9$\msun). Small differences in the black hole
mass will not affect our results, however, 
as the black hole mass is only $\sim$2\%\ of the total mass at
the effective centre of the innermost \chandra\ bin.}.

To fit the three-dimensional entropy profile we assumed a model comprising a 
broken 
powerlaw, plus a constant; such a model is reasonably successful at reproducing the 
entropy profiles of galaxies and galaxy groups over a wide radial range 
\citep{humphrey09d,jetha07a,finoguenov07a,gastaldello07b,sun08a}. In order to 
provide more flexibility in fitting the data (and weight more heavily the outer data-points
in the fit), we also allow an additional break
in the entropy profile at large radius.
The normalization of the powerlaw and constant components, the radius of the breaks
and the slopes above and below them were allowed to fit freely. 

Following \citet{humphrey08a}, we solved the equation of hydrostatic equilibrium to determine
the gas properties as a function of radius, from 10~pc to a large radius outside the 
field of view. 
For the latter, we adopted twice the virial radius of the system defined by ignoring the
baryonic components (which is slightly smaller than the true virial radius of the 
system), but explore alternative choices in \S~\ref{sect_syserr_3d}.
For any arbitrary set of mass and entropy model parameters, it is not always possible to 
find a physical solution to this equation over the full radial range. Such 
models were therefore rejected as unphysical during parameter space exploration. 
In order to compare to the observed {\em projected} density and temperature profiles, we 
projected the three-dimensional gas density and temperature, using a procedure
similar to that described in \citet{gastaldello07a}\footnote{In computing the plasma emissivity 
term, we approximated the true three dimensional abundance profile with the projected abundance profile
(Fig~\ref{fig_kt_rho}); since the projected and deprojected profiles do not differ
significantly, this should be sufficient for our present purposes, but we explore
this question in more detail in \S~\ref{sect_syserr_emissivity}}.
This involves computing an emission-weighted projected mean temperature and density in each 
radial bin, that can be compared directly to the data\footnote{We note that, for systems with
a strong temperature gradient and  kT \gtsim 3~keV the emission-weighted temperature
may be biased low \citep{mazzotta04a}. Nevertheless, we find little evidence of any 
strong bias when we fit the deprojected temperature and density data, which should be 
much less sensitive to this effect. This suggests that the impact of this effect is not
significant here (\S~\ref{sect_syserr_3d}).}.

To compare the model to the data, we used the $\chi^2$ statistic, fitting simultaneously the 
\chandra\ and \suzaku\ temperature and density profiles, with the central bin of the 
\suzaku\ data excluded from our fits. Correlations between the density errors were simply 
implemented by adopting a form for the $\chi^2$ statistic which incorporates the covariance
matrix \citep[\eg][]{gould03a}\footnote{By default we only consider correlations between
the density data, but we investigate introducing a more complete covariance computation in 
\S~\ref{sect_syserr_covariance} and find that the results are not significantly affected.}.
Parameter space was explored with a Bayesian Monte Carlo method. Specifically, we used 
version 2.7 of the MultiNest code\footnote{{http://www.mrao.cam.ac.uk/software/multinest/}} \citep{feroz08a,feroz09a}.
Since the choice of priors is nontrivial, we followed convention in cycling through a 
selection of different priors and assessing their impact on our results 
(we discuss this in detail in \S~\ref{sect_syserr_priors}). Initially, 
however, we adopted flat priors on
the logarithm of the dark matter mass (over the range $10^{12}< M < 10^{16}$\msun), 
the logarithm of the dark matter halo concentration, $c_{DM}$ (over the range $1<c_{DM}<100$),
the logarithm of the gas density at the canonical radius, the stellar M/\lk\ ratio, and the various entropy 
parameters. For a more detailed description of the modelling
procedure, see \citet{humphrey09d,buote11a}. 

\subsection{Mass profile} \label{sect_mass_profile}
\begin{figure}
\centering
\includegraphics[height=3.3in,angle=270]{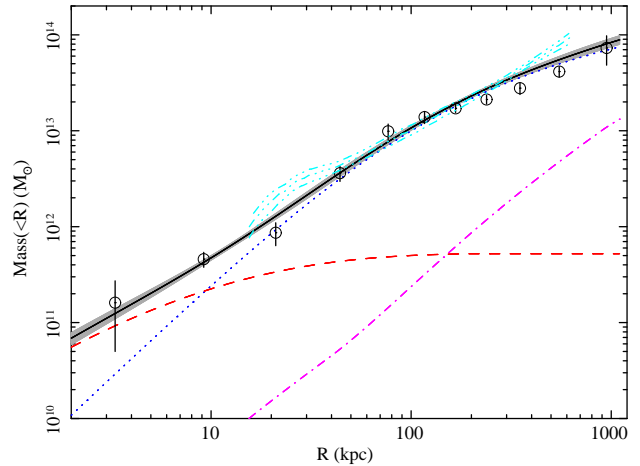}

\caption{Radial mass profile of \src. The solid (black) line
indicates the total enclosed mass (and the grey shaded region
indicates the 1-$\sigma$ error in the total mass distribution),
the dashed (red) line indicates the 
stellar mass, the dotted (blue) line is the dark matter, and the 
dash-dot (magenta) line is the gas mass contribution. 
Overlaid are a set of data-points derived from a more traditional
analysis method described in \citet{humphrey09d}. We stress that the
model is {\em not} fitted to these data, but is derived independently
from the temperature and density data; nevertheless, the agreement
between the different approaches is good. We also show the 
mass profile found by \citet{vikhlinin06b} from the \chandra\
data (light blue dash-dot-dot line), which overall agrees with
our results, although is higher at both large and small radii.
\label{fig_mass_profile}}
\end{figure}
\begin{deluxetable*}{llllllll}
\tablecaption{Mass results and error budget\label{table_mass}}
\tabletypesize{\scriptsize}
%\rotate
\centering
\tablehead{
\colhead{Test} & \colhead{M$_*$/\lk} & \colhead{log \mtwentyfive} & \colhead{log $c_{2500}$}& \colhead{log \mfive} & \colhead{log $c_{500}$} & \colhead{log \mvir} & \colhead{log \cvir} \\ 
& \colhead{\msun\lsun$^{-1}$}& \colhead{[\msun]} & & \colhead{[\msun]} & & \colhead{[\msun]}  
}
\startdata
Marginalized & $0.54 \pm 0.10$& $13.49^{+0.03}_{-0.02}$& $0.44 \pm 0.07$& $13.77 \pm 0.04$& $0.76 \pm 0.06$& $13.97^{+0.05}_{-0.04}$& $1.05 \pm 0.06$\\
 Best-fit& $(0.51)$& $(13.49)$& $(0.48)$& $(13.75)$& $(0.80)$& $(13.95)$& $(1.09)$\\\hline
$\Delta$DM profile&  $+0.19$  $\left( ^{+0.08}_{-0.11}\right)$  &  $-0.024$  $\left( ^{+0.02}_{-0.03}\right)$  &  \ldots &  $+0.06$  $\left( ^{+0.02}_{-0.03}\right)$  &  \ldots &  $+0.21$  $\left( \pm {0.02}\right)$  &  \ldots \\
$\Delta$AC&  $-0.209$  $\left( ^{+0.10}_{-0.06}\right)$  &  $+0.003$  $\left( \pm {0.03}\right)$  &  $-0.015$  $\left( ^{+0.07}_{-0.08}\right)$  &  $+0.008$  $\left( \pm {0.04}\right)$  &  $-0.013$  $\left( ^{+0.06}_{-0.08}\right)$  &  $+0.02$  $\left( ^{+0.04}_{-0.05}\right)$  &  $-0.017$  $\left( \pm {0.07}\right)$ \\
$\Delta$Background&  $^{+0.02}_{-0.06}$  $\left( ^{+0.10}_{-0.13}\right)$  &  $^{+0.01}_{-0.02}$  $\left( \pm {0.03}\right)$  &  $^{+0.02}_{-0.01}$  $\left( \pm {0.07}\right)$  &  $^{+0.01}_{-0.03}$  $\left( \pm {0.04}\right)$  &  $^{+0.02}_{-0.01}$  $\left( \pm {0.07}\right)$  &  $^{+0.02}_{-0.02}$  $\left( \pm {0.04}\right)$  &  $^{+0.03}_{-0.01}$  $\left( \pm {0.07}\right)$ \\
$\Delta$SWCX&  $-0.193$  $\left( ^{+0.11}_{-0.14}\right)$  &  $-0.027$  $\left( \pm {0.04}\right)$  &  $+0.10$  $\left( ^{+0.07}_{-0.09}\right)$  &  $-0.052$  $\left( \pm {0.05}\right)$  &  $+0.09$  $\left( ^{+0.06}_{-0.08}\right)$  &  $-0.047$  $\left( \pm {0.05}\right)$  &  $+0.09$  $\left( ^{+0.06}_{-0.08}\right)$ \\
$\Delta$Stray light&  $-0.005$  $\left( \pm {0.10}\right)$  &  $+0.01$  $\left( ^{+0.02}_{-0.03}\right)$  &  $-0.006$  $\left( \pm {0.07}\right)$  &  $+0.003$  $\left( \pm {0.04}\right)$  &  $-0.006$  $\left( \pm {0.06}\right)$  &  $+0.006$  $\left( \pm {0.04}\right)$  &  $-0.007$  $\left( \pm {0.06}\right)$ \\
$\Delta$Rmax&  $^{+0.02}_{-0.01}$  $\left( \pm {0.10}\right)$  &  $^{+0.01}_{-0.00}$  $\left( \pm {0.03}\right)$  &  $^{+0.01}_{-0.00}$  $\left( \pm {0.07}\right)$  &  $-0.018$  $\left( ^{+0.05}_{-0.03}\right)$  &  $\pm 0.01$  $\left( \pm {0.06}\right)$  &  $\pm 0.01$  $\left( \pm {0.04}\right)$  &  $\pm 0.01$  $\left( \pm {0.06}\right)$ \\
$\Delta$3d &  $+0.05$  $\left( \pm {0.09}\right)$  &  $-0.049$  $\left( \pm {0.04}\right)$  &  $+0.02$  $\left( \pm {0.08}\right)$  &  $-0.053$  $\left( \pm {0.05}\right)$  &  $+0.02$  $\left( \pm {0.07}\right)$  &  $-0.040$  $\left( \pm {0.05}\right)$  &  $+0.02$  $\left( \pm {0.07}\right)$ \\
$\Delta$Fit priors&  $^{+0.03}_{-0.01}$  $\left( \pm {0.11}\right)$  &  $+0.01$  $\left( \pm {0.03}\right)$  &  $^{+0.01}_{-0.03}$  $\left( \pm {0.08}\right)$  &  $\pm 0.01$  $\left( \pm {0.04}\right)$  &  $^{+0.01}_{-0.03}$  $\left( \pm {0.07}\right)$  &  $^{+0.02}_{-0.00}$  $\left( \pm {0.05}\right)$  &  $^{+0.01}_{-0.03}$  $\left( \pm {0.07}\right)$ \\
$\Delta$Stars &  $-0.031$  $\left( ^{+0.00}_{0.00}\right)$  &  $\pm 0$  $\left( ^{+0.02}_{-0.03}\right)$  &  $+0.08$  $\left( \pm {0.10}\right)$  &  $-0.020$  $\left( ^{+0.03}_{-0.05}\right)$  &  $+0.07$  $\left( \pm {0.09}\right)$  &  $-0.022$  $\left( ^{+0.04}_{-0.06}\right)$  &  $+0.05$  $\left( ^{+0.10}_{-0.07}\right)$ \\
$\Delta$Weighting&  $+0.09$  $\left( ^{+0.06}_{-0.09}\right)$  &  $+0.03$  $\left( \pm {0.03}\right)$  &  $-0.082$  $\left( \pm {0.07}\right)$  &  $+0.04$  $\left( \pm {0.04}\right)$  &  $-0.074$  $\left( \pm {0.06}\right)$  &  $+0.05$  $\left( \pm {0.04}\right)$  &  $-0.069$  $\left( \pm {0.06}\right)$ \\
$\Delta$PSF&  $\pm 0.01$  $\left( \pm {0.11}\right)$  &  $+0.01$  $\left( ^{+0.02}_{-0.03}\right)$  &  $-0.012$  $\left( \pm {0.07}\right)$  &  $+0.01$  $\left( \pm {0.05}\right)$  &  $-0.011$  $\left( \pm {0.06}\right)$  &  $+0.02$  $\left( \pm {0.05}\right)$  &  $-0.012$  $\left( \pm {0.06}\right)$ \\
$\Delta$Instrument&  $^{+0.02}_{-0.03}$  $\left( \pm {0.11}\right)$  &  $+0.03$  $\left( \pm {0.03}\right)$  &  $^{+0.02}_{-0.05}$  $\left( \pm {0.08}\right)$  &  $^{+0.04}_{-0.01}$  $\left( \pm {0.05}\right)$  &  $^{+0.02}_{-0.05}$  $\left( \pm {0.07}\right)$  &  $^{+0.04}_{-0.01}$  $\left( \pm {0.05}\right)$  &  $^{+0.02}_{-0.05}$  $\left( \pm {0.07}\right)$ \\
$\Delta$Spectral&  $+0.05$  $\left( \pm {0.10}\right)$  &  $+0.03$  $\left( ^{+0.02}_{-0.03}\right)$  &  $-0.019$  $\left( \pm {0.07}\right)$  &  $+0.02$  $\left( \pm {0.04}\right)$  &  $-0.018$  $\left( \pm {0.06}\right)$  &  $+0.02$  $\left( \pm {0.04}\right)$  &  $-0.016$  $\left( \pm {0.06}\right)$ \\
$\Delta$Distance &  $^{+0.15}_{-0.12}$  $\left( \pm {0.13}\right)$  &  $^{+0.01}_{-0.03}$  $\left( ^{+0.04}_{-0.02}\right)$  &  $\pm 0.11$  $\left( \pm {0.07}\right)$  &  $^{+0.03}_{-0.08}$  $\left( ^{+0.05}_{-0.03}\right)$  &  $\pm 0.10$  $\left( \pm {0.07}\right)$  &  $^{+0.07}_{-0.08}$  $\left( \pm {0.04}\right)$  &  $\pm 0.09$  $\left( \pm {0.06}\right)$ \\
$\Delta$Entropy &  $+0.01$  $\left( ^{+0.08}_{-0.12}\right)$  &  $+0.005$  $\left( \pm {0.03}\right)$  &  $-0.007$  $\left( \pm {0.07}\right)$  &  $-0.003$  $\left( \pm {0.04}\right)$  &  $-0.007$  $\left( \pm {0.06}\right)$  &  $+0.006$  $\left( \pm {0.04}\right)$  &  $-0.006$  $\left( \pm {0.06}\right)$ \\
$\Delta$Covariance &  $^{+0.02}_{-0.03}$  $\left( ^{+0.08}_{-0.15}\right)$  &  $^{+0.00}_{-0.01}$  $\left( \pm {0.03}\right)$  &  $+0.04$  $\left( ^{+0.06}_{-0.08}\right)$  &  $-0.028$  $\left( \pm {0.04}\right)$  &  $+0.03$  $\left( ^{+0.06}_{-0.08}\right)$  &  $-0.017$  $\left( \pm {0.04}\right)$  &  $+0.03$  $\left( \pm {0.06}\right)$ \\
\enddata
\tablecomments{{Marginalized values and 1-$\sigma$ confidence regions for the stellar 
mass-to-light (M$_*$/\lk) ratio and the enclosed mass and concentration measured at various overdensities. 
Since the best-fitting parameters need not be identical to the marginalized
values, we also list the best-fitting values for each parameter (in parentheses). }
In addition to the statistical errors, we also show estimates of the error budget
from possible sources of systematic uncertainty.
We consider a range of different
systematic effects, which are described in detail in \S~\ref{sect_syserr}; 
specifically we evaluate the effect of the
choice of dark matter halo model ($\Delta$DM), adiabatic contraction ($\Delta$AC),
treatment of the background ($\Delta$Background) and the Solar wind charge exchange
X-ray component ($\Delta$SWCX), stray light ($\Delta$Stray light)
maximum radius used in the projection calculation (${\rm \Delta R_{max}}$), 
deprojection ($\Delta$3d), priors on the model parameters 
($\Delta$Fit priors), treatment of the stellar light ($\Delta$Stars),
removing the emissivity correction ($\Delta$Weighting), 
the effect of errors in our treatment of spectral mixing due to the
PSF of  \suzaku\ ($\Delta$PSF), the X-ray detectors used 
($\Delta$Instrument), spectral fitting choices ($\Delta$Spectral),
distance uncertainties ($\Delta$Distance), the parameterization of the 
entropy model ($\Delta$Entropy),
and covariance between the 
temperature and density data-points ($\Delta$Covariance).  
We list the change
in the marginalized value of each parameter for every test and, in parentheses,
the statistical uncertainty on the parameter determined from the test.
Note that the systematic error estimates should {\em not}
in general be added in quadrature with the statistical error.}
\end{deluxetable*}

%Mention that ignoring above 2.0 keV is more dramatic, and introduces a larger systematic error
%(\fgvir is $+0.11$).

We found the model could fit the density and temperature data well, with a best-fitting
$\chi^2$/dof=14.4/15, even when taking into account the covariance between adjacent
density data-points (see \citetalias{humphrey11a}).
The best-fitting models are overlaid in Fig~\ref{fig_kt_rho}. 
In Fig~\ref{fig_kt_rho},
we also show the range of possible {\em three-dimensional} temperature and density
profiles predicted by our model that are consistent with the projected data.
In Table~\ref{table_mass}, we tabulate the best-fitting values and marginalized 
confidence regions for each mass parameter of interest.
The best-fitting radial mass distribution is shown in Fig~\ref{fig_mass_profile},
along with the relative contributions of each of the different mass model components.
Overlaid are a series of mass data points derived from fitting the
 {\em deprojected} data using the ``traditional smoothed inversion'' mass modelling method 
\citep{buote11a}, which is more subject to systematic uncertainties
\citep[for more details see \S~\ref{sect_syserr_3d} and][]{humphrey09d}.
The agreement is very good, indicating that the resulting mass profile is not overly sensitive
to the analysis method.

We obtain a marginalized \rvir=$1100\pm31$~kpc, which compares to an outer \suzaku\
annulus spanning 820--1190~kpc (thus having an ``effective centre'' at $\sim$1000~kpc).
Therefore, we confirm that we are able to reach \rvir\ with these data.
\begin{figure}[!h]
\centering
\includegraphics[width=3.3in]{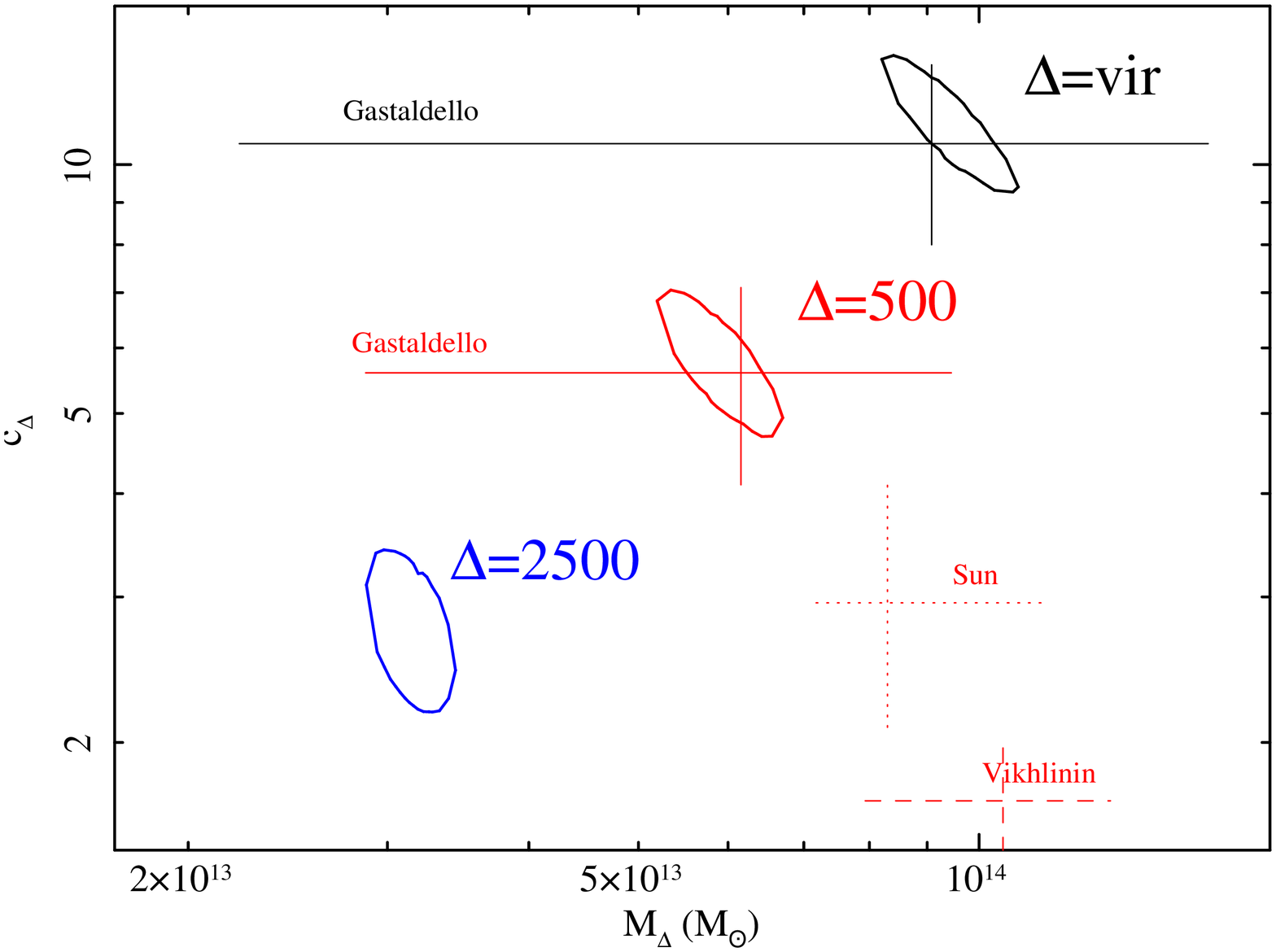}
\caption{Derived relation between mass and concentration for \src\ (contours).
We show the results for several over-densities (virial$\equiv$108, in black; 
500 in red; 2500 in blue). We also compare our results to constraints reported
in the literature \citep{gastaldello07a,vikhlinin06b,sun08a}.\label{fig_cm}}
\end{figure}
\begin{figure*}
\centering
\includegraphics[width=6.5in]{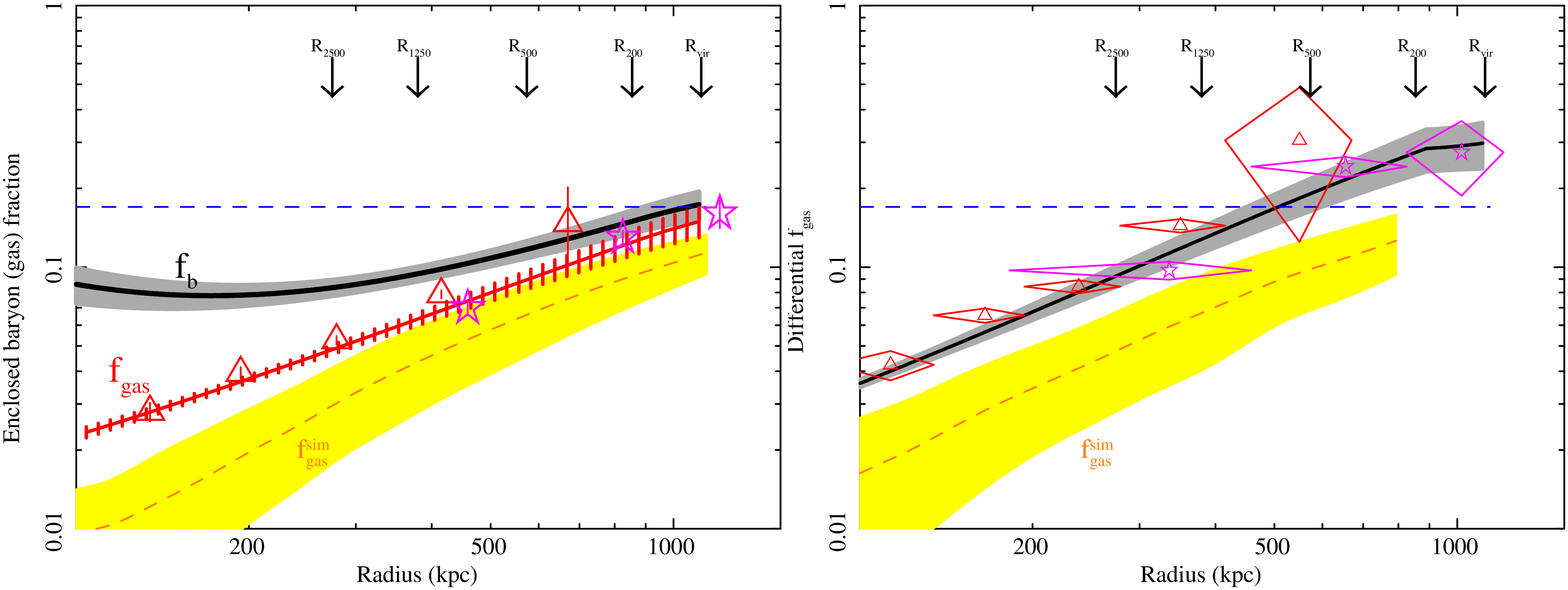}
\caption{{\em Left:} Radial distribution of the enclosed gas fraction (\fgas) and baryon fraction
(\fb) in \src. The blue dotted line indicates the 
Univeral baryon fraction \citep[0.17:][]{dunkley09a,komatsu11a}, while we also indicate various
interesting radii. In yellow, we show the predicted \fgas\ profile at this mass scale
from recent numerical simulations \citep{young11a}. {\em Right:} The {\em local}
(differential) gas fraction profile. At large radii, this exceeds the Universal value,
indicating that gas has been pushed out to these scales.
\label{fig_fgas}}
\end{figure*}
 In Fig~\ref{fig_cm}, we show the relation between the 
concentration of the gravitating mass, \cvir,  and the virial mass, \mvir. 
To be consistent with our past work
\citep{buote07a}, the \mvir\ and \rvir\ are derived from the distribution
of the {\em total} gravitating mass, not just the dark matter, and the concentration,
\cvir\ is defined as the ratio of \rvir\ to the characteristic scale of the DM halo.
We compare our results with concentration and 
mass constraints for \src\ reported in the literature (and based only on the 
\chandra\ data) at different overdensities. 
We find excellent agreement with the \mvir-\cvir\ results from \citet{gastaldello07a};
in particular the agreement with the virial mass and concentration obtained from that
paper is interesting, given that the \citet{gastaldello07a} results at this overdensity
were significantly extrapolated. We do, however, find 
a systematic offset at \rfive\ from the work of \citet{sun08a} and \citet{vikhlinin06b},
who employed essentially the same  ``smoothed inversion'' mass modelling approach 
\citep{buote11a} in each work, and obtained slightly higher masses and lower concentrations
than our best-fitting values. We  discuss the possible origin of these
discrepancies in \S~\ref{sect_discussion_mass}.
\begin{deluxetable*}{llllrrr}
\tablecaption{Baryon fraction results and error budget\label{table_fb}}
\tabletypesize{\scriptsize}
\centering
\tablehead{
\colhead{Test} & \colhead{$f_{g,2500}$} & \colhead{$f_{g,500}$} & \colhead{$f_{g,vir}$}& \colhead{$f_{b,2500}$} & \colhead{$f_{b,500}$} & \colhead{$f_{b,vir}$}\\
}
\startdata
Marginalized & $0.048 \pm 0.001$& $0.087 \pm 0.005$& $0.14 \pm 0.02$& $0.12^{+0.009}_{-0.01}$& $0.124^{+0.007}_{-0.008}$& $0.17 \pm 0.02$\\
Best-fit& $(0.048)$& $(0.089)$& $(0.149)$& $(0.117)$& $(0.127)$& $(0.174)$\\\hline
$\Delta$DM profile&  $+0.002$  $\left( \pm {0.001}\right)$  &  $-0.005$  $\left( \pm {0.004}\right)$  &  $-0.044$  $\left( \pm {0.01}\right)$  &  $+0.03$  $\left( \pm {0.01}\right)$  &  $\pm 0$  $\left( ^{+0.006}_{-0.008}\right)$  &  $-0.046$  $\left( \pm {0.01}\right)$ \\
$\Delta$AC&  $\pm 0$  $\left( ^{+0.002}_{-0.001}\right)$  &  $\pm 0$  $\left( \pm {0.005}\right)$  &  $-0.003$  $\left( \pm {0.02}\right)$  &  $-0.022$  $\left( \pm {0.01}\right)$  &  $-0.012$  $\left( ^{+0.006}_{-0.007}\right)$  &  $-0.009$  $\left( \pm {0.02}\right)$ \\
$\Delta$Background&  $+0.001$  $\left( \pm {0.001}\right)$  &  $^{+0.004}_{-0.001}$  $\left( \pm {0.006}\right)$  &  $^{+0.02}_{-0.01}$  $\left( \pm {0.02}\right)$  &  $^{+0.00}_{-0.00}$  $\left( \pm {0.01}\right)$  &  $\pm 0.00$  $\left( \pm {0.01}\right)$  &  $^{+0.02}_{-0.01}$  $\left( \pm {0.02}\right)$ \\
$\Delta$SWCX&  $+0.001$  $\left( ^{+0.001}_{-0.002}\right)$  &  $+0.003$  $\left( \pm {0.005}\right)$  &  $+0.02$  $\left( ^{+0.01}_{-0.02}\right)$  &  $-0.021$  $\left( ^{+0.02}_{-0.01}\right)$  &  $-0.005$  $\left( ^{+0.01}_{-0.01}\right)$  &  $+0.02$  $\left( ^{+0.01}_{-0.02}\right)$ \\
$\Delta$Stray light&  $\pm 0$  $\left( \pm {0.001}\right)$  &  $-0.002$  $\left( \pm {0.005}\right)$  &  $-0.007$  $\left( \pm {0.02}\right)$  &  $-0.003$  $\left( \pm {0.01}\right)$  &  $-0.003$  $\left( \pm {0.01}\right)$  &  $-0.005$  $\left( \pm {0.02}\right)$ \\
$\Delta$Rmax&  $\pm 0$  $\left( \pm {0.001}\right)$  &  $+0.002$  $\left( ^{+0.005}_{-0.006}\right)$  &  $+0.004$  $\left( \pm {0.02}\right)$  &  $-0.001$  $\left( \pm {0.01}\right)$  &  $-0.001$  $\left( \pm {0.01}\right)$  &  $+0.009$  $\left( ^{+0.01}_{-0.02}\right)$ \\
$\Delta$3d &  $+0.004$  $\left( \pm {0.002}\right)$  &  $+0.01$  $\left( ^{+0.007}_{-0.005}\right)$  &  $+0.03$  $\left( \pm {0.02}\right)$  &  $+0.02$  $\left( \pm {0.01}\right)$  &  $+0.02$  $\left( ^{+0.01}_{-0.01}\right)$  &  $+0.04$  $\left( \pm {0.02}\right)$ \\
$\Delta$Fit priors&  $\pm 0$  $\left( \pm {0.002}\right)$  &  $-0.002$  $\left( \pm {0.005}\right)$  &  $^{+0.003}_{-0.005}$  $\left( \pm {0.02}\right)$  &  $^{+0.00}_{-0.00}$  $\left( \pm {0.01}\right)$  &  $^{+0.00}_{-0.00}$  $\left( \pm {0.01}\right)$  &  $^{+0.00}_{-0.00}$  $\left( \pm {0.02}\right)$ \\
$\Delta$Stars &  $\pm 0$  $\left( ^{+0.001}_{-0.001}\right)$  &  $+0.002$  $\left( \pm {0.005}\right)$  &  $+0.009$  $\left( \pm {0.02}\right)$  &  $^{+0.03}_{-0.04}$  $\left( \pm {0.01}\right)$  &  $\pm 0.02$  $\left( \pm {0.01}\right)$  &  $\pm 0.01$  $\left( \pm {0.02}\right)$ \\
$\Delta$Weighting&  $\pm 0$  $\left( ^{+0.002}_{-0.001}\right)$  &  $-0.002$  $\left( \pm {0.005}\right)$  &  $-0.015$  $\left( ^{+0.01}_{-0.02}\right)$  &  $+0.003$  $\left( \pm {0.01}\right)$  &  $-0.003$  $\left( \pm {0.01}\right)$  &  $-0.016$  $\left( \pm {0.02}\right)$ \\
$\Delta$PSF&  $\pm 0$  $\left( \pm {0.002}\right)$  &  $-0.002$  $\left( \pm {0.005}\right)$  &  $-0.010$  $\left( \pm {0.02}\right)$  &  $-0.005$  $\left( \pm {0.01}\right)$  &  $-0.003$  $\left( \pm {0.01}\right)$  &  $-0.006$  $\left( \pm {0.02}\right)$ \\
$\Delta$Instrument&  $+0.002$  $\left( ^{+0.002}_{-0.002}\right)$  &  $-0.003$  $\left( \pm {0.01}\right)$  &  $-0.028$  $\left( ^{+0.02}_{-0.03}\right)$  &  $-0.006$  $\left( \pm {0.01}\right)$  &  $-0.005$  $\left( \pm {0.01}\right)$  &  $-0.026$  $\left( ^{+0.02}_{-0.03}\right)$ \\
$\Delta$Spectral&  $-0.003$  $\left( \pm {0.001}\right)$  &  $-0.007$  $\left( \pm {0.005}\right)$  &  $-0.020$  $\left( \pm {0.02}\right)$  &  $-0.004$  $\left( \pm {0.01}\right)$  &  $^{+0.00}_{-0.01}$  $\left( \pm {0.01}\right)$  &  $-0.017$  $\left( \pm {0.02}\right)$ \\
$\Delta$Distance &  $^{+0.01}_{-0.01}$  $\left( \pm {0.002}\right)$  &  $^{+0.02}_{-0.01}$  $\left( \pm {0.01}\right)$  &  $^{+0.03}_{-0.02}$  $\left( \pm {0.02}\right)$  &  $+0.003$  $\left( \pm {0.01}\right)$  &  $^{+0.01}_{-0.00}$  $\left( \pm {0.01}\right)$  &  $^{+0.02}_{-0.01}$  $\left( \pm {0.02}\right)$ \\
$\Delta$Entropy &  $\pm 0$  $\left( \pm {0.002}\right)$  &  $\pm 0$  $\left( \pm {0.005}\right)$  &  $-0.001$  $\left( \pm {0.02}\right)$  &  $-0.003$  $\left( \pm {0.01}\right)$  &  $\pm 0$  $\left( \pm {0.01}\right)$  &  $\pm 0$  $\left( \pm {0.02}\right)$ \\
$\Delta$Covariance &  $\pm 0$  $\left( ^{+0.002}_{-0.001}\right)$  &  $+0.004$  $\left( ^{+0.005}_{-0.006}\right)$  &  $^{+0.011}_{-0.002}$  $\left( \pm {0.02}\right)$  &  $-0.002$  $\left( ^{+0.01}_{-0.01}\right)$  &  $^{+0.00}_{-0.00}$  $\left( \pm {0.01}\right)$  &  $+0.01$  $\left( \pm {0.02}\right)$ \\
\enddata
\tablecomments{Marginalized values and 1-$\sigma$ confidence regions for the gas 
fraction ($f_{g,\Delta}$) and baryon fraction ($f_{b,\Delta}$) measured at various overdensities
($\Delta$). We also provide the best-fitting parameters in parentheses, and a breakdown 
of possible sources of systematic uncertainty, following Table~\ref{table_mass}. We find that
$f_b$ is reasonably robust to most sources of systematic uncertainty, especially within 
\rtwentyfive.}
\end{deluxetable*}
\subsection{Gas and baryon fraction constraints}
\begin{figure}
\centering
\includegraphics[height=3.3in,angle=270]{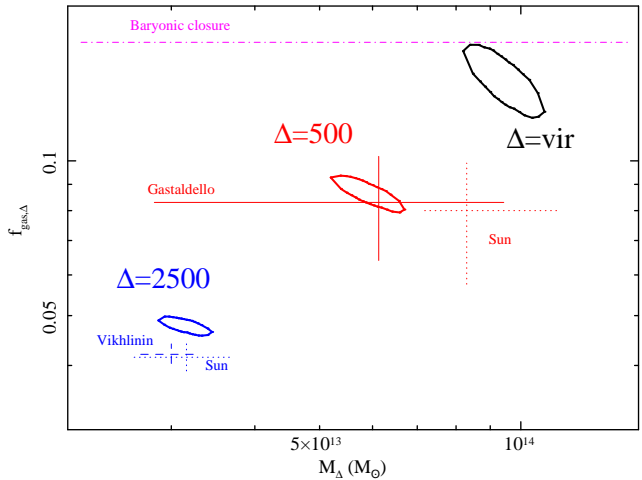}
\caption{Contours of \fgas\ {\em versus} mass at different over-densities.
The magenta (dash-dot) line indicates the Universal baryon fraction (annotated ``baryonic
closure''). We overlay results from the literature. \label{fig_fgas_compare}}
\end{figure}
In Table~\ref{table_fb}, we tabulate the best-fitting values and marginalized
confidence regions for the gas and baryon fractions of the system at different
overdensities. In performing this calculation, we included the mass in baryons
of the central galaxy and the hot gas (inferred from our fit), and also folded in
two additional, uncounted reservoirs of baryons, intra-cluster light and
additional galaxies in the group. \citet{vikhlinin99a} found that $\sim$25\%\
of the V-band stellar light associated with galaxies was in non-central
galaxies. We assumed that this also holds in the K-band, and adopted a 
K-band M/L ratio of 1 for these galaxies. 
Furthermore, for the virial mass of the system, we expect as 
much as $\sim$2 times the stellar mass of the central galaxy will be bound up
in intra-cluster light \citep{purcell07a}. If we make the reasonable assumption
that these uncounted baryons are distributed in the same way as the dark matter
(this is approximately true of the additional galaxy light reported by 
\citeauthor{vikhlinin99a}), the total mass inferred from our modelling
(which did not explicitly include these components) will be correct. 
We discuss the sensitivity of our results to our treatment of these
components in \S~\ref{sect_syserr_stars}.  In practice, however,
these uncounted components account for only $\sim$10\% of the baryon budget, since
most of the baryons are tied up in the hot gas halo. 

In Fig~\ref{fig_fgas}, we display the radial dependence of enclosed 
baryon fraction, which rises modestly
from \fb$\sim$0.1 at 100~kpc to $0.17\pm0.02$ by \rvir, in excellent agreement
with the Cosmological baryon fraction \citep[0.17:][]{dunkley09a,komatsu11a}.
{\em We stress there is no extrapolation in this measurement of \fb\ at the 
virial radius, since the \suzaku\ data reach the virial radius in this system}.
This behaviour is strikingly similar to the 
isolated Galaxy NGC\thin 720 \citepalias{humphrey11a}, 
which has a virial mass $\sim$2 orders
of magnitude smaller than \src.

%Information from Vikhlinin's paper, using the correct cosmology:
%LV (central galaxy) = 1.7e11
%LV (285 kpc) = 2.14e11
%LV (500 kpc) = 2.24e11
%LV (725 kpc) = 2.35e11

% For best-fitting NFW model => total excess LV is 
% 7.6e10

% Additionally, you should add in the ICL component, which is 
% distributed like the stars... from Purcell et al, we expect
% ~50--70% of the BCG+ICL light to be in the ICL; therefore we 
% expect some 1--2.3 times as much ICL as there is light in the 
% central system. Ie the total V-band of the ICL should be 
% 4.1e11 Lsun;

% If this all has the same distribution as the DM; then since the 
% DM is inferred from fitting the (true) gravitating mass,
% Mtot is correct; DM is just ``uncounted matter''....

In Fig~\ref{fig_fgas}, we also show the radial distribution of the enclosed
gas fraction, which rises steeply with radius (as is typically observed 
in groups and clusters, \eg\ \citealt{vikhlinin06b}). For comparison,
we overlay the predictions of recent numerical simulations 
\citep{young11a}, which systematically under-estimate the true gas 
fraction. At small scales
(or higher overdensities) the low \fgas\ indicates that gas has
either been bound up into stars or ``pushed out'' to large 
radii by feedback. However, the approximate baryonic closure of the system
suggests that little gas has been evacuated completely from the system in
such a process.
That gas has been ``pushed out'' in this way is reflected in the differential
gas fraction (\ie the gas density divided by the total mass density at a given
radius), which actually
exceeds the Cosmological baryon fraction outside $\sim$500~kpc (Fig~\ref{fig_fgas},
right panel). In all,
{we find that $\sim$65\%\ of the gas
within \rvir\ actually lies outside \rfive, illustrating the importance of the \suzaku\ data
for directly measuring it.} 

In Fig~\ref{fig_fgas_compare}, we compare our enclosed \fgas\ constraints at different overdensities
with values reported in the literature. 
Once again, our measurements agree well with \citet{gastaldello07a},
but there are some modest discrepancies with \citet{sun08a} and \citet{vikhlinin06b}, who
found slightly smller \fgas. We will discuss the possible origin of these differences in
\S~\ref{sect_discussion_mass}.

\subsection{Entropy profile} \label{sect_entropy}
\begin{figure}
\centering
\includegraphics[width=3.3in]{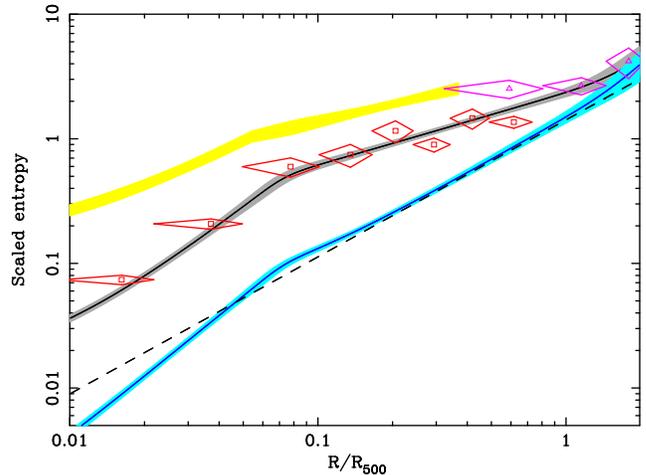}
\caption{Entropy profile model of \src, scaled by its characteristic entropy,
and shown as a function of \rfive. The grey shaded region is the 1-$\sigma$ confidence region
determined from our mass model fit.  We overlay deprojected data-points from \chandra\
(triangles) and \suzaku\ (stars), which are determined more directly from the data, and
agree well with the model. We stress the model is {\em not} fitted to these data. 
We note that the deprojection procedure in large spatial bins likely introduces 
non-neglible (unphysical) noise into the data-points (see \S~\ref{sect_syserr_3d}), 
and so they should be treated with caution. The dotted line indicates
the ``baseline'' predictions from gravitational structure formation \citep{voit05b},
and we find that 
the ``\fgas-corrected'' entropy profile (blue shaded region; see text), which agrees 
well with the baseline predictions out to $\sim$\rvir. 
Also shown (yellow region) is the scaled entropy profile of the isolated elliptical galaxy
NGC\thin 720, illustrating more entropy injection in the lower-mass system.
\label{fig_entropy}}
\end{figure}
As expected for approximately hydrostatic gas, we find that we obtain a good fit to the data
with a model requiring a monotonically rising entropy (S) profile. We show the model profile in
Fig~\ref{fig_entropy} (grey shaded region), scaled by the ``characteristic entropy'' ${K_{500}}$
\citep{voit05b},
and shown as a function of fraction of \rfive\ reached. 
In the inner part of the system, the slope ($S\sim R^{\beta}$) is slightly steeper than
 the canonical $\beta\sim$1.1, and the normalization is significantly enhanced
over the ``baseline'' model for gravity-only structure formation simulations \citep{voit05b}, 
indicating significant entropy injection. Above $\sim$40~kpc ($\sim$0.07\rfive), 
the entropy profile flattens 
significantly, and converges with the (steeper) baseline model by \rvir. For maximum
flexibility, we allowed an additional break at large radii, but found that it was
poorly constrained, and the overall shape of the distribution is very flat 
from $\sim$40~kpc to $\sim$\rvir; in fact, fits omitting the break are able to fit the 
data just as well, with only a minimal impact on the recovered gas properties (\S~\ref{sect_syserr_entropy}).

To provide a less model-dependent view of the entropy profile, we overlay in Fig~\ref{fig_entropy}
a series of data-points, which are directly computed from the deprojected density and 
temperature profiles. These were obtained by emulating the behaviour of the ``projct'' \xspec\
model, and correcting for emission projected into the line of sight from outside the 
outermost annulus (see \citetalias{humphrey11a}). These data agree well with the smooth model,
giving us confidence in our treatment of the data.

Following \citet{pratt10a}, we investigated the effect of scaling the entropy profile by a simple
correction factor, $S_{corr}=S_{obs} (f_g/0.17)^{2/3}$, where $S_{corr}$ is the corrected
entropy profile, and $S_{obs}$ is the observed (scaled) distribution. As for more
massive galaxy clusters (\citeauthor{pratt10a}), and the isolated elliptical galaxy 
NGC\thin 720 \citepalias{humphrey11a}, we find that $S_{corr}$ is in much better agreement with
the baseline entropy model (Fig~\ref{fig_entropy}).

\subsection{\rosat\ surface brightness profile} \label{sect_rosat_sb}
\begin{figure}
\centering
\includegraphics[height=3.3in,angle=270]{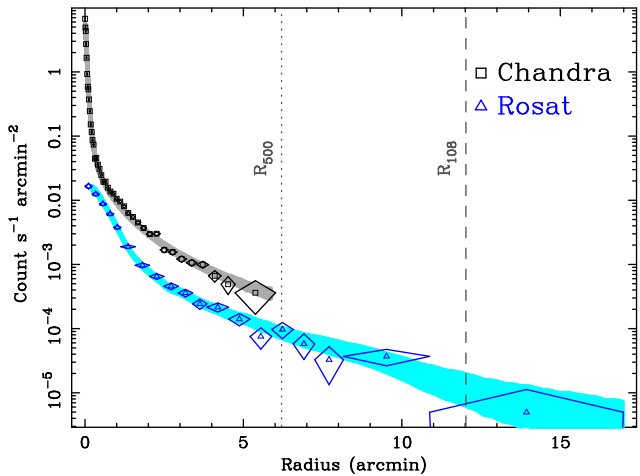}
\caption{Comparison of the flat-fielded, background-subtracted 0.3--7.0~keV \chandra\ (black squares) and 
0.42--2~keV \rosat\ (blue triangles) surface brightness profiles for \src. Overlaid are the profiles
predicted by the best-fitting model; the shaded regions correspond to the 1-$\sigma$ uncertainty.
The vertical displacement between the \rosat\ and \chandra\ data reflects
differences in the effective area curves and energy-bands. In practice, our 
model agrees very well with the \rosat\ data, out as far as $\sim$17\arcmin. 
\label{fig_sb} The vertical lines indicate \rfive\ and \rvir.}
\end{figure}
%BKD paper: \citep{plucinsky93a}.
We next explored whether the model fitted to the \chandra\ and 
\suzaku\ data are consistent with the \rosat\ 
surface brightness profile 
\citep[\eg][]{eckert11a}. To do this, we computed the 
three-dimensional gas emissivity, based on our best-fitting models 
for the temperature, density and abundance profiles. This model was 
projected onto the sky and folded through 
the appropriate \rosat\ PSPC instrumental responses. To account for 
possible mis-calibration between the satellites, we allowed an
arbitrary scaling of the model normalization between \chandra\ 
and \rosat. 
We broadened the surface brightness model by folding in the 
instrumental PSF, evaluated at 1~keV.
We added a constant (sky) background component and 
allowed its normalization, and the aforementioned scaling factor,
to fit to the \rosat\ surface brightness profile, using 
dedicated software 
based around the \minuit\footnote{http://lcgapp.cern.ch/project/cls/work-packages/mathlibs/minuit/index.html} fitting library. 
%Specifically,
%we computed a three-dimensional
%gas emissivity model for \src, based on our best-fitting models for the 
%temperature, density and abundance profiles. This model was projected
%onto the sky and folded through 
%the appropriate \rosat\ PSPC instrumental responses. To account for 
%possible mis-calibration between the satellites, we allowed the 
%normalization of the model to be fitted freely. 
The best-fitting value of the scaling factor ($0.93\pm 0.03$) 
indicates good overall agreement, although there 
may be a modest calibration discrepancy, at least when observing a 
$\sim$1~keV source at large ($\sim$17\arcmin) off-axis angles with \rosat.
Nevertheless, such a modest discrepancy will not affect our conclusions
in the outer part of the group. 
%An additional, constant surface 
%brightness component, 
%was allowed to fit freely to account for the cosmic X-ray background 
%and Galactic foreground components (the instrumental background was 
%subtracted; see \S~\ref{sect_rosat}). We broadened the surface model by folding in the 
%instrumental PSF, evaluated at 1~keV. 
The surface brightness profile was 
fitted out to $\sim$25\arcmin, and became consistent with the background
outside $\sim$12\arcmin. 

In Fig~\ref{fig_sb}, we show the background-subtracted \rosat\ surface
brightness profile (triangles) and, for comparison, the \chandra\ ACIS-S3
data. We overlay the predicted surface brightness models, illustrating 
excellent agreement with the \rosat\ data out beyond \rvir. This strongly
supports our treatment of the \suzaku\ (and \chandra) data, and 
confirms that the entropy profile does not exhibit substantial flattening
outside $\sim$\rfive.

\section{The cosmic X-ray background} \label{sect_cxb}
\begin{figure}
\centering
\includegraphics[height=3.3in,angle=270]{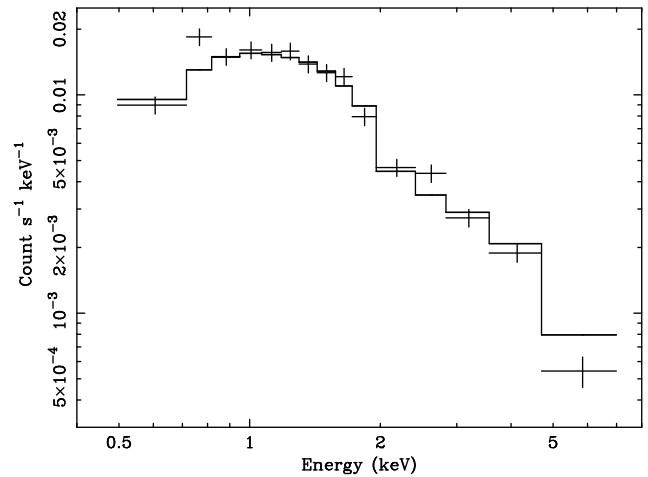}
\caption{Composite spectrum of the detected \chandra\ point-sources in the 
region of interest with fluxes $<2\times 10^{14}$\ergpscm. The brightest few
sources were omitted so that they do not unduly bias the fit. Overlaid is the 
canonical powerlaw ($\Gamma=1.41$) used to fit the CXB component, which
gives a reasonably good overall fit to the data. 
\label{fig_cxb_spectrum}}
\end{figure}
\begin{figure}
\centering
\includegraphics[height=3.3in,angle=270]{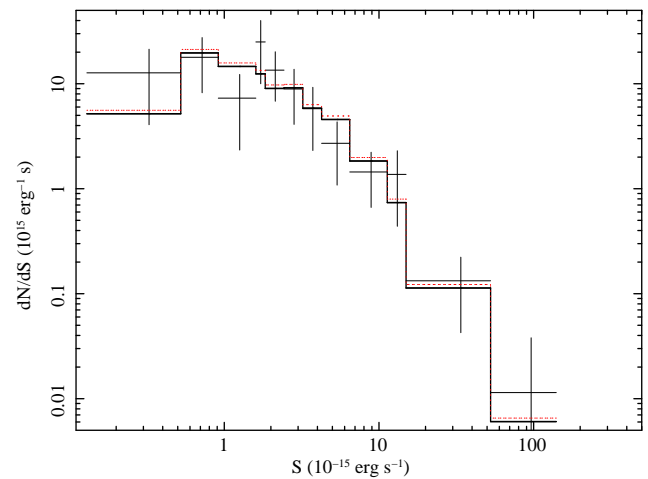}
\caption{The measured differential log N-log S relation for point-sources
within the \chandra\ field of view. The down-turn at low fluxes is due to 
source detection incompleteness; we fold incompleteness and Eddington bias
corrections into the fitted model. The dotted red line is the expected
fit, based on the hard-source counts in \citet{luo08a}, and the 
solid black line is the CDF-S best-fitting model, rescaled to fit the data.
\label{fig_cxb_xlf}}
\end{figure}
As is clear from Fig~\ref{fig_spectra}, the accurate determination of the 
gas properties in the outer two \suzaku\ annuli requires the background to
be determined with high accuracy (see also \S~\ref{sect_syserr_background}).
The dominant background component of relevance is actually the cosmic 
X-ray background (CXB) resulting from (unrelated) undetected, background
point-sources. In this section, we explore the accuracy with which the 
CXB component has been fitted in our \suzaku\ analysis.

\chandra's spatial resolution allows a significant fraction of the
CXB to be resolved into individual point sources, at least near the 
optical axis. By disentangling them from any diffuse 
emission, resolving the sources in this way allows the CXB spectral shape
and normalization to be determined more precisely. 
Deep \chandra\ observations
have yielded accurate measurements of the logN-logS distribution (and hence,
average surface brightness) of
background point sources along particular sightlines \citep[\eg][]{luo08a}.
Alternatively, X-ray spectra from regions of the sky free of foreground
contamination have allowed the CXB shape and normalization to be 
carefully calibrated, averaged over small portions of the sky
\citep[\eg][]{deluca04a}.
However,  due to cosmic variance and stochastic effects, the inferred 
surface brightness will not be well-enough known 
to use this information along other, arbitrary sightlines, such as 
that to \src. 
Ideally, therefore \chandra\ should be used to resolve the point-sources
in each of the  \suzaku\ annuli, and the resultant 
spectra can be used to refine the \suzaku\ analysis.
Unfortunately, the more obstructed field of view of \chandra\ 
(at least in the ACIS-S configuration), coupled with the substantial 
degradation of the PSF (and hence, reduced detection efficiency) 
at \gtsim 4\arcmin\ away
from the optical axis, means that multiple \chandra\ observations are
needed to mosaic the entirety of each \suzaku\ annulus at high enough
precision. To complicate matters further, 
since point sources can be variable, contemporaneous observations 
with each satellite would ideally be used. Nevertheless,
while individual sources can vary significantly, the integrated 
source properties are not expected to be strongly affected by 
variability \citep[\eg][]{kraft01,zezas04a}.

Since the existing \chandra\ data did not meet these requirements, it 
was not possible to improve our \suzaku\ constraints using this approach.
Nevertheless, it is still important to verify {\em consistency} between
the \chandra\ and \suzaku\ CXB measurement.
To do this, we first examined the sources detected by \chandra\ that 
lay within each \suzaku\ region.
%\footnote{For simplicity, we focused
%on the XIS0 regions. Formally, the field of view of each XIS unit does
%not exactly correspond to the same region, on account of the exclusion of 
%calibration sources. Nevertheless, in fitting the \suzaku\ data, we 
%normalized all spectra to match XIS0, and ignored this subtlty.}.
In the outermost region, we detected  9 sources, which we estimate
(below) to contribute only $\sim$20\%\ of the flux within this aperture.
This estimate reflects both the fact that only $\sim$50\%\ of this 
region is exposed with \chandra, and the large off-axis angles 
($\sim$9-13\arcmin) under scrutiny. In the 2--5\arcmin\ aperture, however,
the \chandra\ data were more helpful; we estimate that $\sim$80\%\ of the CXB
flux was resolved into 26 detected sources. These estimates, of course,
assume that the detected sources did not vary in brightness significantly between the 
\chandra\ and \suzaku\ observations. 

In principle, one can sum the spectra of all detected sources within each
region, and use that to determine the shape (if not the normalization) of the 
CXB spectrum. However, this estimate suffers from a bias, since
it assumes that the spectrum of a background AGN is 
independent of its flux, which is not true. If we omitted sources above a 
particular flux limit from this calculation, the composite spectrum was 
systematically flatter. Fits to the CXB spectrum measured from
unresolved background emission in source-free regions of the sky 
should not be affected by this problem, and so in our default analysis, we 
parameterized the CXB spectrum as a powerlaw with $\Gamma=1.41$, as found
by \citet{deluca04a}. The incompleteness correction of the composite spectrum 
that is necessary to test formal consistency between
this model and the \chandra\ data is beyond the scope of this paper. 
However, in Fig \ref{fig_cxb_spectrum}
we show the composite spectrum of sources fainter than $2\times 10^{-14}$\ergpscm\
(omitting the few, bright point sources in the field so as not to skew unfairly
the spectral shape), which agrees reasonably well (if not perfectly) with this
model. In \S~\ref{sect_syserr_cxb} we find that modest variation in the slope of the
CXB component ($\pm$5\%) does not strongly affect our conclusions.

While the CXB shape may vary modestly between sightlines, the uncertainty on
its normalization is likely to be more problematical for our analysis. To verify
consistency between the \suzaku\ flux in each aperture and the \chandra\ 
esimates requires an estimate not only of the mean flux, but also the 
uncertainty on it, from undetected sources in each region. 
To measure the mean flux, it was first necessary to
determine the normalization and shape of the average logN-logS distribution
for the point sources along the \src\ line of sight. Since cosmic variance is unlikely
to be strong between each \suzaku\ region, 
we considered all  sources detected by \chandra\ that lay within the \suzaku\ XIS0 field of 
view. Following \citet{humphrey08b}, we 
subtracted a local background from each source spectrum and converted the 
counts to 0.5--7.0~keV flux, assuming a powerlaw model ($\Gamma$=1.55) with Galactic 
absorption, while taking into account the spatial dependence of the effective area.
Given the measured fluxes and spatial distribution, we do not believe any of these sources 
to be X-ray binaries associated with the central galaxy.
The resulting logN-logS distribution (in $dN/dS$ form) is shown in Fig~\ref{fig_cxb_xlf}. 
In order to interpret these data, it was necessary to fit a model,
correcting for both incompleteness at low fluxes and the Eddington bias.
In \citet{humphrey08b}, we discuss in detail how these corrections
were carried out. In short, we corrected the model ({\em not} the data)
based on the results of extensive Monte Carlo simulations in which
fake sources were added (in a spatially uniform, random fashion)
to the \chandra\ images, and the source detection algorithm was used to try to 
detect them and determine their flux.

We found that the $dN/dS$ distribution could be well-fitted with a 
(incompleteness-corrected)
broken powerlaw model (with a break at $7.4\times 10^{-15}$\ergpscm, and 
negative logarithmic differential slopes 1.3 and 2.6 below and above
the break, respectively)
that we also found could fit the hard-band logN-logS distribution
reported by \citet{luo08a} for the \chandra\ Deep Field South (CDF-S) line of 
sight\footnote{Taking into account the different energy-bands 
used, and the spectral models used to convert counts to flux.}. 
The best-fitting model is shown in Fig~\ref{fig_cxb_xlf}, which is very
close to the fit to the CDF-S data (shown, correctly normalized
for the observed region of sky). This good agreement gives us confidence in our
treatment of the data.

\begin{figure}
\centering
\includegraphics[height=3.3in,angle=270]{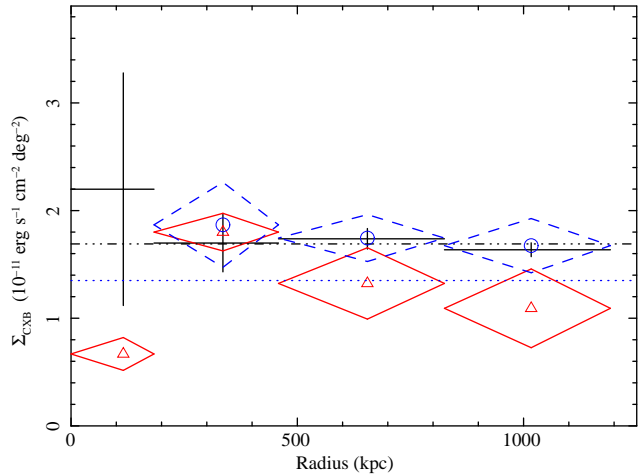}
\caption{The 0.5--7.0~keV surface brightness profile ($\Sigma_{CXB}$) of the CXB component, measured
from the \suzaku, \rosat\ and \chandra\ data. We show the best-fitting $\Sigma_{CXB}$
from the default \suzaku\ analysis (black dash-dot-dot line), and $\Sigma_{CXB}$ determined
independently from the \suzaku\ data in each region (black crosses). Also shown are
the estimates from the \rosat\ surface brightness profile (dashed error bars; 
blue circles) and from fitting 
the \chandra\ point source distributions (red triangles). The blue dotted line is the 
expected flux from integrating the logN-logS point source distribution along the CDF-S 
line-of-sight \citep[][see text]{luo08a}. We find excellent agreement between the 
different estimates. Note that the CXB component need not be constant 
with radius, and so we only expect good agreement only between the fitted data-points, not the 
approximate (constant) models shown. 
\label{fig_cxb_norm}}
\end{figure}
While the average normalization of the CXB component along the \src\ 
sightline can be determined in this way, the actual flux measured in
any annulus is subject to stochastic scatter about this value due to
the discrete source nature of the CXB. To account for this, we undertook
Monte Carlo simulations, as outlined below:

{\noindent\hangindent 10pt 1. Using the best-fitting logN-logS relation, described
above, we determined the total number of sources per square degree, N$_0$ with
flux $>10^{-16}$\ergpscm.  On
each simulation, we added Gaussian noise to N$_0$ to reflect uncertainties ($\pm 12$\%)
in the fit to the logN-logS relation.}

{\noindent\hangindent 10pt 2. For each \suzaku\ region $i$, which has area $A_i$, 
we drew N$_i$ sources
that have fluxes distributed like the logN-logS relation between $10^{-16}$ and
$10^{-12}$\ergpscm. N$_i$ was itself Poisson distributed around $N_0 A_i$.}

{\noindent\hangindent 10pt 3. Based on the fraction of the \suzaku\ region covered
by \chandra, and our source detection incompleteness estimates at each flux, every source has a 
particular probability of being detected. Using this information, we randomly assigned
a state (detected, or not detected) to each source.}

{\noindent\hangindent 10pt 4. In each region $i$, we estimated the CXB flux by 
summing up the flux of all simulated
sources flagged as ``undetected'', and added in the flux of the {\em real} detected sources
and the expected flux of sources below $10^{-16}$\ergpscm\ (by integrating the 
logN-logS relation). }

{\noindent\hangindent 10pt 5. We repeated steps 1--4 a large number of times to 
determine the average CXB flux and its uncertainty in each region $i$. This estimate
accounts for both stochastic scatter and the uncertainty on the logN-logS normalization
from our fit.}

\noindent In Fig~\ref{fig_cxb_norm}, we show the estimated X-ray flux for each bin.
along with the value measured from the \suzaku\ spectral fitting. We show both the 
best-fitting value for our default analysis (in which the CXB normalization per
square degree is tied between all the annuli), and the constraints on the CXB flux
when the the normalization is allowed to fit freely in each region. The \chandra\
results appear to be in good agreement with the \suzaku\ measurements\footnote{We note that 
the slight difference in the powerlaw slopes used to flux the sources ($\Gamma=1.55$)
and to fit the CXB ($\Gamma=1.41$) leads the \chandra\ data to underestimate the \suzaku\
flux only by $\sim$6\%, which is far smaller than the statistical errors.}.
If the integrated flux is very sensitive to the properties of individual,
bright point sources, we would expect the CXB surface brightness measured by \suzaku\ to 
show significant scatter between each radial bin, in excess of the measured
statistical error. In fact, the CXB surface brightness is consistent with
being constant with radius.

As a final consistency check, we also explored whether the \rosat\ data agreed  
with the \suzaku\ results. We extracted the \rosat\ radial surface brightness
profile in the band 1--2~keV; based on the best-fitting \suzaku\ background model,
we expect that $\sim$95\%\ of the background emission in this region comes from the unresolved CXB. 
We restricted the photons to be from the region of sky imaged by XIS0 
(although we also excluded a 3.4\arcmin-wide region in the vicinity of the 
inner support ring shadow). We subtracted off the particle background, the 
expected (small) contribution from the soft Galactic background components (which was
assumed to be spatially uniform), and the surface brightness profile
predicted from our best fitting mass model (see \S~\ref{sect_rosat_sb}). The remaining
counts should come from the CXB emission. Rebinning the profile to match the \suzaku\
annuli, and converting counts to flux by folding the canonical 
CXB model through the \rosat\ response matrices (computed close to the centre of \src),
we obtained an estimate of the CXB surface brightness in excellent agreement with the 
\suzaku\ measurements (Fig~\ref{fig_cxb_norm}).

\section{Systematic error budget} \label{sect_syserr}
%\label{sect_syserr_covariance} \label{sect_syserr_3d} \label{sect_syserr_priors} 
In this section, we address the sensitivity of our results to
various data analysis choices that were made.
In most cases, it is difficult or impractical to
express these assumptions through a single additional model
parameter over which one might hope to marginalize, and so we
adopted the pragmatic approach of investigating how our results
changed if the assumptions were arbitrarily adjusted. We focused
on those systematic effects likely to have the greatest impact on
our conclusions, and list in Tables~\ref{table_mass} and \ref{table_fb}
the change to the marginalized value of each key parameter for each
test.  We outline below how each test was performed. 
%Those
%readers uninterested in the technical details of our analysis may
%wish to proceed directly to Section~\ref{sect_discussion}

\subsection{Dark Matter profile} \label{sect_syserr_mass} 
One of the major sources of uncertainty on the recovered mass model 
is the coice of DM mass model. While the NFW model is theoretically
motivated, we also experiemented with the 
so-called ``cored logarithmic'' mass model \citep{binney08a}. This model
tends to predict higher masses (by $\sim$60\%), and correspondingly smaller 
gas fractions, at large scales. Although we cannot distinguish between
the NFW and cored logarithmic models on the basis of $\chi^2$ alone,
the ratio of the Bayesian evidence ($2.6\times 10^{-3}$) implies that the 
cored logarithmic model, with the adopted priors 
(a flat prior on the 
asymptotic circular velocity, between 10 and 2000~$km\ s^{-1}$, and a flat 
prior on $log_{10} r_c$, where $r_c$ is the core radius, over the range
$0\le log_{10} r_c \le 3$.) is a poorer description of the data at $\sim$3.0-$\sigma$.

Another modification to the DM halo profile that is well-motivated
theoretically, although less secure observationally
\citep{humphrey06a,gnedin07a,napolitano10a}, is so-called
``adiabatic contraction''  \citep{blumenthal86a,gnedin04a,abadi09a},
where the DM halo density profile reacts to the gravitational influence of 
baryons that are condensing into stars by becoming cuspier. Modifying the 
NFW profile with the algorithm of 
\citet{gnedin04a}\footnote{Using the CONTRA code publicly available from http://www.astro.lsa.umich.edu/$\sim$ognedin/contra/}, has only a very slight
effect on the best-fitting mass model (``$\Delta$AC'' in Tables~\ref{table_mass} and \ref{table_fb}). This reflects that the scale radius of the DM 
halo is much larger than the effective radius of the stellar mass 
component.

\subsection{Background} \label{sect_syserr_background}\label{sect_syserr_cxb} 
Since the data were background-dominated in the outermost \suzaku\ 
annuli, the treatment of the background was a potentially serious
source of systematic uncertainty. To investigate the extent to
which our results are sensitive to this, we explored a range of 
different choices. First, for the \chandra\ data, we adopted the 
standard blank-field events files distributed with the CALDB to
extract a background spectrum for each annulus.
Since the blank-field  files for each CCD have different 
exposures, spectra were accumulated for each CCD individually, scaled to a
common exposure time and then added. The spectra were renormalized 
to match the observed count-rate in the 9--12~keV band. These ``template'' spectra were then used as a background in \xspec, and the background model components were
omitted from our fit. This gave a formally poorer fit, but did  not 
strongly affect our conclusions. 
Second, since the \suzaku\ non X-ray background
spectra generated by {\em xisnxbgen} may be uncertain, we allowed their
normalization to scale by $\pm$5\%.  

At the temperature of the gas in the outermost annuli ($\sim$1~keV), the
dominant background component is the CXB. In \S~\ref{sect_cxb}, we 
demonstrated that the best-fitting CXB model to our data is in excellent
agreement with predictions for the line of sight to \src. However,
there were still uncertainties in our treatment. Specifically, by default,
we assumed a constant surface brightness for the CXB component, which
may not be formally correct. We therefore experimented with allowing the 
CXB normalization to fit freely in each \suzaku\ annulus 
(Fig~\ref{fig_cxb_norm}). This did not significantly affect our results. 
Additionally, there is uncertainty on the spectral shape of the CXB component.
Given the statistical uncertainty on the shape of the CXB parameterization
found by \citet{deluca04a}, we varied the slope of the CXB powerlaw 
component by $\pm 5$\%. Although this affected the error on the gas 
density at the largest radii, our conclusions were not significantly 
altered. Allowing a larger change in the slope did have a more significant
effect on the density and temperature, however, motivating the need for
deep \chandra\ data to resolve and constrain the CXB at the largest 
scales (\S~\ref{sect_discussion_cxb}).
We summarize the results in Tables~\ref{table_mass} and \ref{table_fb} 
(``$\Delta$Background'').

\subsubsection{Solar Wind Charge Exchange} \label{sect_syserr_swcx}
An additional background component can arise from 
the interaction of the Solar wind with interstellar material
and the Earth's exosphere. This should manifest itself as a 
time-variable, soft component that can be modelled as a series
of narrow Gaussian lines, the intensity of which correlate 
with the Solar wind activity \citep[\eg][]{snowden04a}.  
We did not explicitly include components to account for this 
``Solar Wind Charge Exchange'' (SWCX) in our fits, 
although the soft background components are partially degenerate
with it. To explore whether the SWCX could have affected our conclusions, 
we used the Solar Wind Ion Composition Spectrometer (SWICS)
instrument aboard the Advanced Composition Explorer (ACE) 
spacecraft 
\citep{mccomas98a}\footnote{Based on the 
publicly released data available
from http://www.srl.caltech.edu/ACE/ASC/index.html} to 
identify periods of enhanced SWCX emission. 
Following \citet{snowden04a}, we assumed it to be negligible 
when the ${\rm O^{+7}/O^{+6}}$ ratio falls below $\sim$0.2, and 
selected times during each observation which met this criterion.
While the \chandra\ data were moderately affected ($\sim$60\%\ of 
the data were excluded by this cut), in the low surface brightness
regime, the \suzaku\ data were most important, and these 
were only mildly affected ($\sim$9\%\ of the data were removed
by this cut; the 0.5--1.0~keV count-rate was enhanced only by
$\sim$6\%\ if all the data were used).
The corrected \chandra\ and \suzaku\ spectra were 
fitted to obtain the temperature and density profiles, and folded through
our mass modelling apparatus. We found that our results were not substantially
affected by correcting for the SWCX component (``$\Delta$SWCX'' in 
Tables~\ref{table_mass} and \ref{table_fb}).

%\subsection{Entropy profile parameterization} \label{sect_syserr_entropy}
%The extrapolation of the entropy profile is one of the 
%prime sources of uncertainty in our determination of \fb\ at large scales.
%As we will show in \S~\ref{sect_entropy},
%our default profile is, in fact, consistent with trends seen in massive
%galaxy clusters \citep{pratt10a}, giving us confidence in its use. Nevertheless, 
%it is important to explore plausible alternative shapes. 
%Motivated by observed entropy profiles in galaxy groups and by 
%the monotonically rising entropy profiles required by hydrostatic
%equilibrium, we investigated two pathological extrapolations that should
%bound any plausible model. {Specifically, we adopted a powerlaw shape
%(entropy $\propto R^\gamma$)
%outside $\sim$80~kpc, with $\gamma=0$ (\ie\ a flat entropy profile), which
%is a secure lower boundary assuming stablity against convection, or 
%$\gamma=2.5$. This latter value was chosen arbitrarily, and is far larger
%than has been observed in real systems.}
%This choice primarily affects the extrapolated temperature, rather 
%than the density profile, and so \fb\ does not change substantially
%with these choices (``$\Delta$Entropy'' in Tables~\ref{table_mass} and \ref{table_fb}; 
%Fig~\ref{fig_fb_prob}).
%%DDONE

\subsection{Stray light} \label{sect_syserr_pointsrc}
Stray light from bright point sources within $\sim 1^\circ$ can,
in principle, be scattered by the mirrors into the field of view,
and provide an additional source of background. By far the brightest
point source identified by the Rosat All-Sky Survey (RASS) within
this angular distance from \src\ is is NGC\thin 3998, a foreground 
Sy1 galaxy which is 17\arcmin\ from \src.
To investigate the potential impact of stray light,
we included an additional component in our spectral modelling,
corresponding to the contamination from NGC\thin 3998 to each
annulus. We estimated the amount of stray light leakage by using
{\tt xissimarfgen} to generate suitable ancillary response files
for each annulus, with the locus of NGC\thin 3998 as the origin
for the photons. 
We modelled the emission from NGC\thin 3998 as an absorbed powerlaw,
with $N_H$, $\Gamma$ and the normalization fixed to the values derived
from archival XMM-Newton data by \citet{ptak04a}. We found that the 
stray light contamination was $<20$\%\ of the source flux below
\ltsim 2~keV. The stray light component is much harder than the 
$\sim$1~keV spectrum of the hot gas, and is partially degenerate
with the CXB spectral component. Therefore, we found that 
adding this component had minimal effect on our results
(``$\Delta$Stray light'' in Tables~\ref{table_mass} and \ref{table_fb}).

\subsection{Projection/ Deprojection} \label{sect_syserr_3d}
In our analysis, we modelled the projected temperature and density in 
each annulus by evaluating the hydrostatic model for the temperature
and density in three dimensions, and projecting it onto the line of sight.
In principle, the results can be sensitive to the outermost radius used
in the projection calculation. By default, we adopted 2\rvir, but 
we also explored varying this limit between \rvir\ and 3\rvir. This had
a minor impact on our results (``$\Delta R_{max}$'' in Tables~\ref{table_mass} and
\ref{table_fb}).

In this work, we fitted the projected, rather than the deprojected
data (as done, for example, in \citetalias{humphrey06a}). In general, fitting
the projected data leads to smaller statistical error bars, but potentially larger 
systematic uncertainties \citep[\eg][]{gastaldello07a}, and so it is important to investigate
the likely magnitude of such errors. To do this, we examined the effect on our results
of spherically deprojecting the data. We achieved this by using the 
\xspec\ {\em projct} model\footnote{In practice, it was more convenient
to emulate the behaviour of projct by adding multiple ``vapec'' plasma
models in each annulus, with the relative normalizations tied 
appropriately \citep[\eg][]{kriss83a}. This allowed data from the multiple 
\suzaku\ instruments to be fitted simultaneously.}. To account for emission
projected into the line of sight from regions outside the outermost annulus,
we added an apec plasma model to each annulus, with abundance 0.2 (consistent
with the outermost annulus) and the temperature and normalization determined
from projecting onto the line of sight the best-fitting gas temperature and 
density models described in \S~\ref{sect_method}, but considering the models
only outside $\sim$1200~kpc.
%A deprojected analysis (based on a slightly different treatment of the 
%\chandra\ background) was also presented for this system in 
%\citet{humphrey10a}, and so we also fitted our
%hydrostatic models to the temperature and density profile found in that work. 

We note that the results from a deprojection analysis in the large radial
bins used here should be treated with considerable caution. Since the procedure
assumes constant density and emissivity (hence temperature and abundance)  in each 
shell, which is substantial simplication (see Fig~\ref{fig_kt_rho}), this can
introduce significant, unphysical noise (\eg\ see \S~3.3 of \citealt{buote00c};
\citealt{finoguenov99}). Nevertheless, we found that the best-fitting 
derived results were not significantly affected when the deprojected
(rather than the projected) temperature and density profiles were used 
(``$\Delta$Deprojection'' in 
Tables~\ref{table_mass} and \ref{table_fb}).
In Fig~\ref{fig_mass_profile}, we show
a series of mass ``data-points'' obtained from the deprojected density
and temperature profiles, using the  ``traditional''
mass analysis method described in \citet[][see also 
\citealt{humphrey10a}]{humphrey09d}. These data agree very well with the 
best-fitting mass model found in our projected analysis.
Similarly, the entropy profile derived directly from the 
deprojected data displays overall agreement with the results of our projected
analysis, although the data-points exhibit unphysical ``choppiness''
due to the deprojection noise (Fig~\ref{fig_entropy}).
%DDONE

%\subsection{Spectral fitting} \label{sect_syserr_spectra}
%We investigated a few further sources of systematic uncertainty associated with 
%our spectral analysis. Specifically, we considered the impact of changing the neutral
%hydrogen column density by $\pm 25$\%, using a MEKAL plasma model rather than an APEC
%model, or changing the  

\subsection{Priors} \label{sect_syserr_priors}
Since the choice of priors on the various parameters is arbitrary in our analysis,
it is important to determine to what extent they could affect our conclusions.
To do this, we replaced each arbitrary choice in turn with an alternative, reasonable prior.
Specifically, for each parameter describing the entropy profile, we switched from a 
flat prior on that parameter to a flat prior on its logarithm. 
%We replaced the flat
%prior on the stellar M/L ratio with a Gaussian prior, the mean and $\sigma$ of which
%were determined from the stellar population synthesis model results reported in \citetalias{humphrey06a}. 
We used a flat prior on the DM halo mass, rather than on its logarithm,
 and, instead of the flat prior on $\log c_{DM}$, we adopted the distribution of 
c around M found by either \citet{buote07a} or \citet{maccio08a} as a (Gaussian) prior.
The effect of these choices is no larger than the statistical errors on each parameter,
especially for the baryon fraction measured at $R_{200}$ or higher overdensities 
(``$\Delta$Fit priors'' in Tables~\ref{table_mass} and \ref{table_fb}).
%%DDONE

\subsection{Stellar light} \label{sect_syserr_stars}
Since the scale radius of the DM halo is very much larger than
the effective radius of the stellar light, we would not expect
a careful treatment of the stellar light to be important for
accurately measuring the total mass of the system
\citep[in contrast with the lower-mass, elliptical galaxy,
 regime:][]{humphrey06a}. To illustrate this, we excluded the 
central $\sim$20~kpc (roughly twice \reff) from our analysis. 
Since the stellar mass is relatively unimportant at this 
radial scale (Fig~\ref{fig_mass_profile}), we fixed the stellar
M/L ratio to its best-fitting value. This had a minimal impact 
on our conclusions.

Our measurement of \fb\ includes a canonical amount of 
intra-cluster light, which was not directly detected. 
Since the true contribution of this component (and its M/L ratio) is 
unknown, we explored how significantly \fb\ was affected 
if this component was completely omitted, or if its M/L ratio
was fixed at 1 (slightly higher than the best-fitting M/L ratio
for the central galaxy).
These choices did not substantially affect our results
(``$\Delta$Stars'' in Tables~\ref{table_mass} and \ref{table_fb}).
%DDONE

\subsection{Emissivity correction} \label{sect_syserr_emissivity}
{In our default analysis, the projected temperature and density
profile were weighted by the gas emissivity,
folded through the instrumental responses
\citep[for details, see Appendix B of][]{gastaldello07a}. Since the
computation of the gas emissivity 
assumes that the three dimensional gas abundance profile
is identical to the projected profile (which is unlikely to be true), 
it is important to assess how
sensitive our conclusions are to the emissivity correction.  To do
this, we adopted the extreme approach of ignoring the spatial 
variation of the gas emissivity altogether. We found that this 
had a very small effect on our results
($\Delta$Weighting in Tables~\ref{table_mass} and \ref{table_fb}). 
}

\subsection{PSF correction} \label{sect_syserr_psf}
Our \suzaku\ analysis results depend on the 
careful treatment of spectral mixing between each annulus,
which occurs due to the modest spatial resolution of the 
telescope (see \citetalias{humphrey11a}; \citealt{reiprich09a}). 
Our approach was to calculate mixing between the annuli with the
{\tt xissimarfgen} task, which uses ray-tracing. To explore whether
small errors in this procedure could affect our results, we 
experimented with scaling the amount of light that is scattered 
into each annulus by $\pm$5\%. This did not appreciably affect
our conclusions (``$\Delta$PSF'' in Tables~\ref{table_mass} and
\ref{table_fb}).

\subsection{Remaining tests} \label{sect_syserr_instrument} \label{sect_syserr_covariance} \label{sect_syserr_distance} \label{sect_syserr_radius} \label{sect_syserr_spectra} \label{sect_syserr_entropy}
We here outline the remaining tests we carried out, as summarized
in Tables~\ref{table_mass} and \ref{table_fb}. First of all, since the 
inter-calibration of the \suzaku\ XIS units may not be perfect,
we experimented with using only one of the units in the \suzaku\ analysis,
and cycled through each choice.
We also considered using only the \chandra\ data. We found that our
results were resilient to this choice
(``$\Delta$Instrument'').

To assess the impact of various spectral-fitting data analysis choices on 
our results, in turn we varied the neutral hydrogen column density 
by $\pm$25\%, performed the fit over a restricted energy range (0.7--7.0~keV), 
and replaced the APEC plasma model with a MEKAL model. The impact of 
these choices is comparable to the statistical errors on the parameters
(``$\Delta$Spectral'').
To examine the error associated with distance uncertainties, 
we varied the distance to \src\ by $\pm$30\%, finding the effect, particularly
on \fb\ and \fgas\  to be relatively minor (``$\Delta$Distance'').

In order to provide maximum flexibility in our fitting at large radius,
our parameterization of the entropy profile included an arbitrary break at 
large radius. We found that, in general, the break radius tended to large
scales, consistent with it not being required in our fits. We experimented
with fitting a model without this break, obtaining a good fit 
($\chi^2$/dof=14.6/17) in close agreement with our default model
(``$\Delta$Entropy'').

Finally, to examine the possible errors associated with our 
treatment of the covariance between the density data-points, we investigated 
adopting a more complete treatment that considers the covariance between all the 
temperature and density data-points, as well as adopting the more standard 
(but incorrect) approach of ignoring the covariance altogether. The best-fitting
$\chi^2$ value did not change significantly ($\pm$5) with these different choices,
and so the effect on our fits is not
large (``$\Delta$Covariance'' in Tables~\ref{table_mass} and \ref{table_fb}).
%Note on CXB: $\Gamma=1.41\pm 0.06$ (90\%\ c.l. from \citet{deluca04a}); our +/-5\%\
%is consistent with this.

\section{Discussion} \label{sect_discussion}
By combining our new deep \suzaku\ observation of the very relaxed 
fossil group/ poor cluster \src\ with 
archival \chandra\ data, we have now obtained an unprecedented census of the baryons and 
dark matter over almost 3 orders of magnitude in radial range, from $\sim$kpc scales to
the virial radius ($R_{108}=1100$~kpc). We discuss here the implications of these
new measurements, in detail.

\subsection{Hydrostatic equilibrium}
Our best-fitting hydrostatic model fits the density and temperature data-points
extremely well, strongly supporting the hydrostatic approximation for
this system; despite highly nontrivial temperature and density profiles,
a smooth, physical mass model
and a monotonically rising entropy profile (required for stability
against convection) were able to reproduce them well. This would
require a remarkable conspiracy between the temperature and density
if this approximation were seriously  in error. At the crucial,
largest scales, it is striking that we see no evidence of the peculiarities,
such as a flat entropy profile or very high \fgas\ value, that characterized 
the disturbed systems Perseus and Virgo \citep{simionescu11a,urban11a},
and might point to deviations from the hydrostatic approximation in those
systems. 
This agreement between the models and data holds
even down to $\sim$kpc scales, which had been omitted in some previous
studies \citep{vikhlinin06b,sun08a}. While there is weak evidence for a small
disturbance within the central $\sim$10~kpc (\S~\ref{sect_chandra}), 
recent studies of systems that are manifestly more disturbed than \src\
indicate that the mass derived from azimuthally-averaged temperature and
density profiles is relatively robust to such features 
\citep[global errors being $\sim$10--20\%:][for a detailed discussion]
{churazov08a,buote11a}.
In any case, whether or not we include data at these small scales
has only a minimal effect on our conclusions (\S~\ref{sect_syserr_stars}). 

The closeness of the 
system to hydrostatic equilibrium is unsurprising, 
given its round, relaxed-looking  X-ray isophotes
(Fig~\ref{fig_images}). 
Numerical simulations of structure formation suggest that deviations
from hydrostatic equilibrium in morphologically relaxed-looking 
systems are not large, so that errors in the recovered mass 
distribution  should be no larger than $\sim$25\%\
\citep[\eg][]{tsai94a,buote95a,nagai07a,piffaretti08a,fang09a}. 
Finally, it is important to note that, as a (highly evolved)
fossil group, the accretion rate of the system should be small,
hence possible associated 
morphological and dynamical perturbations at the largest
scales will be minimized.

%\begin{itemize}
%\item Relatively short section here; less controversial in groups/ clusters!
%\item Good fit to complex profiles with monotonically rising entropy profile; remarkable
%conspiracy if seriously in error. The mass profile parameters are physically well-behaved 
%(\S~\ref{sect_discuss_mass}
%\item HE holds even in the central region (\ltsim30~kpc) that were ignored by 
%some previous studies. 
%\item Good agreement unsurprising given the relaxed X-ray morphology; numerical sims suggest
%errors no larger than $\sim$25\%\ \citep[\eg][]{tsai94a,buote95a,nagai07a,piffaretti08a,fang09a}.
%\end{itemize}
\subsection{Mass Profile} \label{sect_discussion_mass}
The total mass profile of \src\ {\em out to the virial radius} 
is very well fitted by a model comprising
an NFW dark matter halo, a stellar component and gas (plus, at the smallest scales, a 
black hole), in good accord with theory. 
We note that such agreement is not guaranteed for any model, since at
least one other model (the cored logarithmic
potential) is rejected at $\sim$3-$\sigma$ in our fitting
(\S~\ref{sect_syserr_mass}). 
 Previous studies have suggested that fossil groups
may have significantly enhanced concentration parameters, likely due
to their early epoch of formation \citep[\eg][]{khosroshahi07a}. 
In the case of \src, however,
the halo concentration parameter ($11.2\pm1.6$) is in 
good agreement both with theoretical predictions (from dark matter-only
simulations) for this mass
range \citep{bullock01a,maccio08a} and with the empirical relation derived
by \citet{buote07a}\footnote{Strictly speaking, this comparison is 
not independent, since \src\ was one of the systems used to
determine the \citet{buote07a} relation, albeit with much poorer
data. Nevertheless, \src\ does not appear to be unusual in
comparison to the other systems used by these authors, 
and so it is unlikely that this relation is largely driven by this
one data-point.}. Similar results were obtained for the other 
fossil systems discussed in \citetalias{gastaldello07a} and
\citet{sun08a}.

Our best-fitting mass model parameters are in excellent agreement
with the \chandra\ measurements made by \citetalias{gastaldello07a},
although with much smaller error bars 
(Fig~\ref{fig_cm}). The reduction in the size of the errors reflects
both the addition of the high-quality \suzaku\ data at large radii
(which helps to pin down the flattening of the mass profile at these
scales), and various recent enhancements made to our mass-modelling
procedure that improve the robustness of the fit to sources of 
systematic uncertainty \citep[see][]{buote11a}. In particular, by
rigorously enforcing the Schwarzschild criterion for stability against
convection, the parameter space is substantially shrunk. 

While our results agree with \citetalias{gastaldello07a}, puzzling
discrepancies exist with the other published mass parameterizations
\citep{vikhlinin06b,sun08a}. These fits used only the \chandra\ data,
and eliminated the central $\sim$40~kpc from the fit. The mass 
distribution was inferred from the gas density and temperature profiles 
using the traditional ``smoothed inversion'' method \citep[][for a 
review]{buote11a}, which can introduce non-negligible systematic
errors \citep[\eg][]{humphrey09d}. In Fig~\ref{fig_kt_rho},
we show the (projected) temperature and (deprojected) density profiles
of \citet{vikhlinin06b}, while we show their recovered mass profile
in Fig~\ref{fig_mass_profile}. While the temperature and density
data agree well with our results, there is modest disagreement
in the mass distribution at large (\gtsim\ 300~kpc) and small (\ltsim\ 50~kpc)
scales. We attribute these discrepancies to systematic errors in 
the smoothed inversion technique, but they cannot explain the significant
differences in the parameterized fits. Fitting an NFW model to these data,
we obtain best-fitting parameters ($\log M_{500}=13.8$, $\log c_{500}=0.8$) that
are very close to our measurements, albeit with large fit residuals at 
large scales. 
%(The profile is actually well fitted by a powerlaw $M\propto r^{3-\alpha}$,
%where $\alpha=1.9$, in contrast with $\alpha=1.67^{+0.11}_{-0.10}$ measured from our data
%within $\sim$100~kpc: \citealt{humphrey10a}.) 
Rather than fitting the 
enclosed mass distribution directly, \citeauthor{vikhlinin06b} actually fitted the gravitating
mass density profile derived from their data. In this case, an
NFW density model fit was primilary constrained by the 
data at large radii, yielding a much lower concentration. In both
cases, however, a simple powerlaw actually fitted the profiles
better than NFW. We therefore attribute the lower concentration
and higher mass found by \citeauthor{vikhlinin06b} to systematic 
errors in their derivation of the density profile, which is inherently
more uncertain than the enclosed mass. Since \citeauthor{sun08a}
did not provide individual mass profiles in their paper, it was
not possible to carry out a similar comparison with their work.

\subsection{The baryon fraction}
\begin{figure}
\centering
\includegraphics[height=3.3in,angle=270]{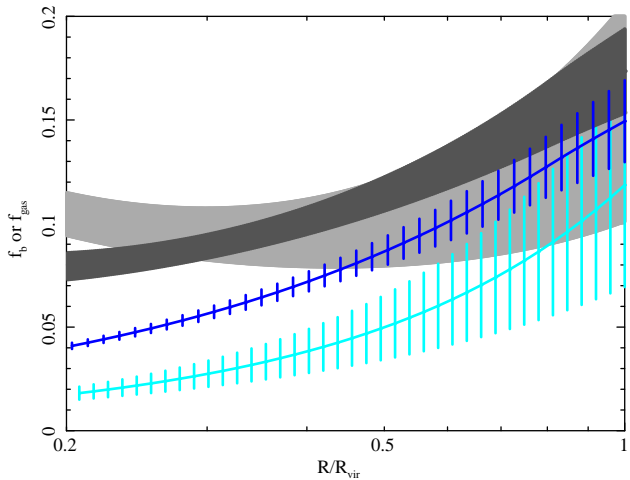}
\caption{The radial profile of \fb\ and \fgas\ for \src\
(dark grey, dark blue regions, respectively), versus fraction of 
\rvir. In comparison, we also show the same quantities for the 
isolated elliptical galaxy NGC\thin 720 (light grey, light blue,
respectively). Note the overall similarity in the \fb\ profile shape,
despite significant discrepancies in \fgas.\label{fig_fb_fgas_rvir}}
\end{figure}
We were able to place tight constraints on both the gas and baryon
fraction out to the virial radius of the system. 
While \fgas\ rises
fairly steeply with radius (so that $\sim$60\%\ of the total
gas mass lies outside $\sim$\rfive), we found that \fb\ rises more 
modestly from $\sim$0.1 at $\sim$100~kpc to approach the 
Universal baryon fraction \citep{dunkley09a,komatsu11a} by \rvir. This is 
consistent with the picture that X-ray bright groups may be 
baryonically closed \citep{mathews05a}. Intriguingly, in the 
isolated, Milky Way-mass elliptical galaxy NGC\thin 720 
(almost $\sim$2 orders of magnitude lower in mass),
we found a remarkably similar 
trend. In Fig~\ref{fig_fb_fgas_rvir}, we compare the profiles 
for the two systems, scaled with respect to the virial radius.
Despite the different \fgas\ profiles, \fb\ is quite consistent
over a wide radial range.
\citet{giodini09a} similarly found a conspiracy between the 
stellar and gas mass to produce a total baryon fraction at \rfive\
that is relatively insensitive to the virial mass of the halo. 
This indicates that, whatever processes are involved in 
redistributing the baryons in the system, in general the baryons
are not being ejected, at least down to $\sim$Milky Way masses.

While \fb\ asymptotes to the Universal value by \rvir, the 
{\em local} gas fraction exceeds this limit outside $\sim$\rfive.
Such an enhancement is to be expected if feedback is redistributing
baryons in the potential well, but not completely evacuating
them from the system. The apparent delicate balance between 
the gravitational potential and the energy injection from feedback
that allows, over a wide range of virial masses, the 
gas to be pushed out almost to \rvir, but not evacuated
from the system, may simply reflect that the mass of gas available
to fuel feedback scales with the mass of the system.

Within \rtwentyfive\ and \rfive, our \fgas\ measurements were in good
agreement with \citetalias{gastaldello07a}, although 
slightly higher than those found by \citet{vikhlinin06b} 
and \citet{sun08a} (see Fig~\ref{fig_fgas_compare}). 
This reflects a slightly lower gas density 
found by those authors (\eg\ Fig~\ref{fig_kt_rho}), at least
within \rtwentyfive. These results are also in good accord with
similar measurements made in other systems 
(see Fig~11 of \citetalias{humphrey11a}), indicating that
\src, despite being a fossil group, does not experience a highly
unusual evolution of its IGrM. In contrast to the outskirts of 
Perseus, we found no evidence of a systematically over estimated
\fgas, which might hint at additional processes such as 
deviations from hydrostatic equilbrium or sphericity, or 
clumpiness of the IGrM \citep{simionescu11a}. The low accretion
rate expected for an old fossil group may reduce the impact of 
these effects in \src. We note that we cannot entirely rule
out clumping in the outer parts of \src, but if present, it must
operate in a finely-balanced conspiracy with \fb\ in such a 
manner as to produce a total baryon fraction very close to the
cosmological value. In our opinion, such a scenario falls 
foul of Occam's razor.

\subsection{Entropy profile} \label{sect_discuss_entropy}
\begin{figure}
\centering
\includegraphics[height=3.3in,angle=270]{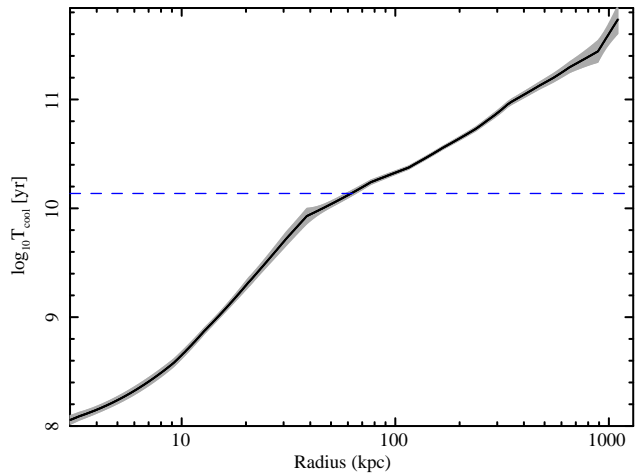}
\caption{Cooling time profile of \src. Most of the gas 
in the system has a cooling time longer than the Hubble
time (indicated with the dashed line). \label{fig_tcool}}
\end{figure}
The entropy profile of \src\ shows remarkably similar characteristics to
that of other galaxy groups \citep[\eg][]{gastaldello07b,mahdavi05a,finoguenov07a,sun08a,cavagnolo09a,johnson09b,flohic11a}. In the innermost regions, the profile rises steeply
(close to $S\propto r^{1.1}$, but offset upwards from the baseline
entropy model, \citealt{voit05b}), then flattens outside $\sim$0.06\rfive\ 
(where the cooling time $\sim 10^{10}$Gyr: Fig~\ref{fig_tcool}), and 
converges with the baseline model by $\sim$\rfive. 
As expected for 
hydrostatic gas, S rises monotonically with radius, and we found little
evidence of any further breaks in the entropy profile.
This overall shape is 
similar to what is predicted for models in which both stellar winds and 
AGN contribute to feedback \citep{mccarthy10a}.
Significantly,
we did not find the characteristic flattening outside $\sim$\rfive\ reported
for Perseus and Virgo \citep{simionescu11a,urban11a}, either in the 
deprojected data, or in our fits with a parameterized model (which had
sufficient parameters to capture such behaviour). Therefore, whatever
processes are involved in producing such a feature cannot be ubiquitous
in massive groups or clusters (or, at least, fossil groups). 

As seen in massive galaxy clusters \citep{pratt10a}, the central parts of 
groups \citep{flohic11a} and the isolated elliptical galaxy NGC\thin 720
\citepalias{humphrey11a}, 
when we scaled the entropy profile by $(f_{gas}/0.17)^{2/3}$,
we found it was brought into fairly close agreement with the baseline,
adiabatic model (Fig~\ref{fig_entropy}).
For \src, we have now confirmed that this relation holds even {\em out to the virial radius}.
Taken together with the local \fgas\ constraints, it seems that the 
primary effect of feedback on the baryons is to redistribute the hot gas 
within the halo, rather than to raise its temperature, or eject it \citep{mathews11a}.

\subsection{The pressure profile of \src}
\begin{figure}
\centering
\includegraphics[height=3.3in,angle=270]{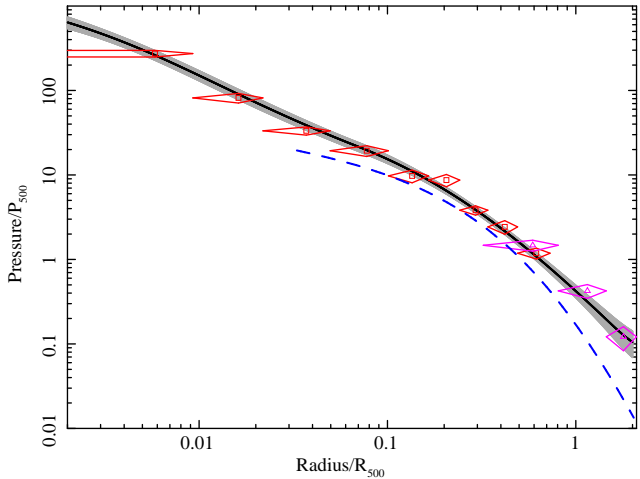}
\caption{Pressure profile for \src. The dashed blue line is the canonical
model of \citet{arnaud10a}.
\label{fig_pressure}}
\end{figure}
Recently, it has become evident that 
the thermal pressure profiles of
galaxy clusters and groups, when scaled appropriately, exhibit remarkable
uniformity, at least between $\sim$0.5--1.0\rfive,
which has important implications for understanding and calibrating standard
scaling relations. 
\citet{arnaud10a} proposed a canonical pressure profile model, based
on observed clusters within $\sim$\rfive\ and numerical simulations
at larger scales. \citet{sun11a} confirmed that this profile holds 
for galaxy groups within $\sim$\rfive. With \src, it is now possible to
test this model at larger scales.  In Fig~\ref{fig_pressure}, we show
the scaled pressure profile of \src, with the canonical model of 
\citeauthor{arnaud10a} overlaid. At small scales (\ltsim 0.1\rfive), 
there is some disagreement, but this is not unexpected, based on 
the cluster and group samples. There is general agreement between
$\sim$0.2--0.5\rfive, but the model significantly diverges from
the observed pressure distribution at larger scales. Since 
the global properties\footnote{\mfive=$(5.9\pm0.5)\times 10^{12}$\msun; 
$kT_{500}=1.78\pm 0.05$, 
where $kT_{500}$ is the emission-weighted temperature within the
projected aperture (0.1--1)\rfive; 
$Y_X=(9.1\pm0.6)\times 10^{12} M_\odot\ keV$,
where $Y_X$ is the gas mass within \rfive\ times $kT_{500}$; see 
\citet{sun08a}.} of \src\
are in agreement with 
standard scaling relations \citep[\eg][]{sun08a}, it is unlikely
that this represents an artefact of, for example, its fossil group
nature. More measurements of the pressure
profile out to $\sim$\rvir\ are clearly needed.
%\begin{itemize}
%\item For comparison with recent work, here's the gas pressure profile out to \rvir.
%\item Disagreement with \citet{arnaud10a} profile, both at small scales and 
%large (\gtsim 0.5\rfive) scales (where simulations dominate the profile). Still,
%scatter is expected from system-to-system, so more constraints are needed before
%strong conclusions drawn.
%\item YX=$(9.1\pm0.6)\times 10^{12} M_\odot\ keV$; 
%kTX=1.78$\pm0.05$ (doesn't matter how you compute it).
%EZ$^{2/5}$=1.03. Compares to YX from Sun=$9.9\times 10^{12}$.
%\end{itemize}

\subsection{Constraining the CXB} \label{sect_discussion_cxb}
Despite the data being background dominated at large radius, our 
results appear to be reasonably robust. This is implied by the well-behaved
nature of the density, temperature and mass profiles, the good
agreement between the different satellites (in particular, the 
agreement with the \rosat\ surface brightness profile), and 
the reliable measurement of the CXB normalization in our \suzaku\
fits. Nevertheless, the gas density at the largest scales was 
sensitive to the precise background, specifically the 
CXB spectrum (which is the dominant background component). There
are, therefore, 
real concerns that, if the adopted model is not accurate,
we may be mis-characterizing the gas properties in the 
outskirts of the group. As discussed in \S~\ref{sect_cxb}, 
one way to mitigate these worries is to employ a \chandra\ 
mosaic of the outer \suzaku\ annuli, enabling us to resolve
a large fraction of the CXB component into point sources,
from which the CXB shape and normalization can be directly
measured. Combining \chandra\ and \suzaku\
in this way exploits their complementary capabilities
(the high spatial resolution of \chandra; the low 
instrumental background and good spectral resolution 
of \suzaku), and should allow the robust detection of 
gas out to \rvir\ to become routine.

%\begin{itemize}
%\item Robust!!
%\item Discuss the importance of constraining the CXB with a deep \chandra\
%mosaic. Also, that deep data may make it possible to measure the hot gas, too,
%although that will be challenging.
%\end{itemize}

\section{Conclusions}
In our joint \suzaku, \chandra\ and \rosat\ analysis of \src, we found:

{\hangindent 10pt 1. By choosing objects with an advantageous 
combination of distance, mass and surface brightness, it is possible to 
detect the gas to \rvir\ in a single \suzaku\ pointing. This approach 
provides better observational efficiency than the multiple pointings
needed to mosaic nearby systems, while providing better azimuthal
coverage.}

{\hangindent 10pt 2. Applying this approach to the 
morphologically relaxed fossil group \src, we were able to 
measure the distribution of gas and gravitating matter out 
to \rvir\ for the first time in a system
with a mass as low as $10^{14}$\msun.}

{\hangindent 10pt 3. Within \rvir, the total
baryon fraction approached the cosmological value, implying
that feedback only redistributed the baryons within the halo,
rather than ejecting them.}

{\hangindent 10pt 4. The entropy profile was enhanced
over the model from gravity-only cosmological simulations, 
implying significant feedback. However, outside $\sim$0.06\rfive,
the profile flattened, and converged with the gravity-only model near
\rvir. After correcting for the gas fraction, the entropy profile was 
close to the self-similar predictions of gravitational structure formation 
simulations, as observed in massive galaxy clusters.}

{\hangindent 10pt 5. In contrast with the Virgo and Perseus
clusters, 
there was no evidence of further flattening in the entropy profile 
outside $\sim$\rfive, nor unusually high \fgas\ measurements.
This indicates that significant gas clumping cannot be ubiquitous
near $\sim$\rvir, at least for relaxed (fossil) groups.}

{\hangindent 10pt 6. There is good agreement between the 
best-fitting model for the gas emission and the \chandra, \suzaku\ and \rosat\ data, 
as well as the agreement in the cosmic X-ray background (the dominant background component)
measurements made with each satellite. }

{\hangindent 10pt 7. The dominant uncertainty in the gas properties at
large radius, the shape CXB spectrum, can be significantly improved upon by 
using a \chandra\ mosaic of the outermost \suzaku\ regions
to resolve the CXB into its component point sources so its properties can be 
constrained directly.
}

%In our joint \suzaku-\chandra\ analysis of \src, we find:
%\begin{itemize}\vspace{-9pt}
%\item Using \suzaku, we were able to constrain the temperature and density of the gas
%out to \rvir.\vspace{-9pt}
%\item At large radius, the gas is well-described by our hydrostatic mass model, 
%implying it is close to hydrostatic.\vspace{-9pt}
%\item We were able to place tight constraints on the concentration and virial mass of the system,
%finding excellent agreement with our previous studies using only \chandra\ 
%\citep{gastaldello07a}.\vspace{-9pt}
%\item We obtained a total gas fraction within \rvir\ of \fgas$=0.16\pm0.01$, consistent
%with baryonic closure. Most ($\sim$65\%) of the gas is found between $\sim$\rfive--\rvir.\vspace{-9pt}
%\item The entropy profile is enhanced over the ``baseline'' prediction for gravity-only 
%structure-formation simulations, but converges with the prediction at $\sim$\rvir.\vspace{-9pt}
%\item When corrected for \fgas, the entropy profile agrees well with the baseline model, even
%out to $\sim$\rvir.\vspace{-9pt}.
%\end{itemize}

%\clearpage
\acknowledgements
We would like to thank Taotao Fang and Norbert Werner for helpful 
discussions. We would also like to thank D.\ Eckert for help with the 
\rosat\ analysis.
This research has made use of data obtained from the High Energy Astrophysics
Science Archive Research Center (HEASARC), provided by NASA's Goddard Space
Flight Center. This research has also made use of the
NASA/IPAC Extragalactic Database (\ned)
which is operated by the Jet Propulsion Laboratory, California Institute of
Technology, under contract with NASA, and the HyperLEDA database
(http://leda.univ-lyon1.fr). We are grateful to the ACE/ SWICS instrument
team for making their data publicly available through the ACE Science Center.
PJH and DAB gratefully acknowledge partial support from NASA under Grant NNX10AD07G,
issued through the office of Space Science Astrophysics Data Program. 
We are also grateful for partial support from NASA-\suzaku\ grant
NNX09AV71G.

\bibliographystyle{apj_hyper}
\bibliography{paper_bibliography.bib}

\end{document}